\documentclass[twocolumn,epjc3]{svjour3}  

\smartqed  
\RequirePackage{fix-cm}
\RequirePackage{graphicx}

\journalname{Eur. Phys. J. C}

\usepackage[utf8]{inputenc}
\usepackage{units}
\usepackage{siunitx}
\usepackage{textgreek}
\usepackage{xcolor}
\usepackage[numbers,sort&compress]{natbib}
\usepackage{bmpsize}
\usepackage{subcaption}
\usepackage{rotating}
\usepackage{multirow}
\usepackage[intlimits]{amsmath}
\usepackage{mathtools}
\usepackage[colorlinks,citecolor=blue,urlcolor=blue,linkcolor=blue,breaklinks=true]{hyperref}
\usepackage{upgreek}
\newcommand{\kgd}{{kg$\cdot$d}}
\newcommand{\ctskgdkeV}{cts\,kg$^{-1}$\,d$^{-1}$\,keV$^{-1}$}
\newcommand{\fluxcms}{cm$^{-2}$\,s$^{-1}$}

\newcommand*{\doi}[1]{\href{https://dx.doi.org/#1}{\nolinkurl{https://dx.doi.org/#1}}}

\title{Simulation-based design study for the passive shielding of the COSINUS dark matter experiment}
\date{Compiled on \today}

\author{G.~Angloher\thanksref{inst1} 
        \and
        I.~Dafinei\thanksref{inst2}
        \and
        N.~Di~Marco\thanksref{te1,inst3,inst4}
        \and
        F.~Ferroni\thanksref{inst2,inst3}
        \and
        S.~Fichtinger\thanksref{inst5}
        \and
        A.~Filipponi\thanksref{inst4,inst6}
        \and
        M.~Friedl\thanksref{inst5}
        \and
        A.~Fuss\thanksref{te2,inst5,inst7}
        \and
        Z.~Ge\thanksref{inst8}
        \and
        M.~Heikinheimo\thanksref{inst9}
        \and
        K.~Huitu\thanksref{inst9}
        \and
        R.~Maji\thanksref{inst5,inst7}
        \and
        M.~Mancuso\thanksref{inst1}
        \and
        L.~Pagnanini\thanksref{inst3,inst4}
        \and
        F.~Petricca\thanksref{inst1}
        \and
        S.~Pirro\thanksref{inst4}
        \and
        F.~Pr\"obst\thanksref{inst1}
        \and
        G.~Profeta\thanksref{inst4,inst6}
        \and
        A.~Puiu\thanksref{inst3,inst4}
        \and
        F.~Reindl\thanksref{inst5,inst7}
        \and
        K.~Sch\"affner\thanksref{inst1}
        \and
        J.~Schieck\thanksref{inst5,inst7}
        \and
        D.~Schmiedmayer\thanksref{inst5,inst7}
        \and
        C.~Schwertner\thanksref{inst5,inst7}
        \and
        M.~Stahlberg\thanksref{inst1}
        \and
        A.~Stendahl\thanksref{inst9}
        \and
        F.~Wagner\thanksref{inst5}
        \and
        S.~Yue\thanksref{inst8}
        \and
        V.~Zema\thanksref{inst1}
        \and
        Y.~Zhu\thanksref{inst8}
        \\(The COSINUS Collaboration)\thanksref{}
        and L.~Pandola\thanksref{LNS}
}

\thankstext{te1}{Corresponding author: \href{mailto:natalia.dimarco@lngs.infn.it}{natalia.dimarco@lngs.infn.it}}
\thankstext{te2}{Corresponding author: \href{mailto:alexander.fuss@oeaw.ac.at}{alexander.fuss@oeaw.ac.at}}

\institute{Max-Planck-Institut f\"ur Physik, 80805 M\"unchen - Germany \label{inst1}
        \and
        INFN - Sezione di Roma, 00185 Roma - Italy \label{inst2}
        \and
        Gran Sasso Science Institute, 67100 L'Aquila - Italy \label{inst3}
        \and
        INFN - Laboratori Nazionali del Gran Sasso, 67010 Assergi - Italy \label{inst4}
        \and
        Institut f\"ur Hochenergiephysik der \"Osterreichischen Akademie der Wissenschaften, 1050 Wien - Austria \label{inst5}
        \and
        Dipartimento di Scienze Fisiche e Chimiche, Universit\`a degli Studi dell'Aquila, 67100 L'Aquila - Italy \label{inst6}
        \and
        Atominstitut, Technische Universit\"at Wien, 1020 Wien - Austria \label{inst7}
        \and
        SICCAS - Shanghai Institute of Ceramics, Shanghai - P.R.China 200050 \label{inst8}
        \and
        Helsinki Institute of Physics, 00560 Helsinki - Finland \label{inst9}
        \and
        INFN - Laboratori Nazionali del Sud, 95125 Catania - Italy \label{LNS}
}

\date{Received: date / Accepted: date}

\begin{document}

\maketitle

\begin{abstract}
    The COSINUS (Cryogenic Observatory for SIgnatures seen in Next-generation Underground Sear\-ches) experiment aims at the detection of dark matter-induced recoils in sodium iodide (NaI) crystals operated as scintillating cryogenic calorimeters. The detection of both scintillation light and phonons allows performing an event-by-event signal to background discrimination, thus enhancing the sensitivity of the experiment.  
    The construction of the experimental facility is foreseen to start by 2021 at the INFN Gran Sasso National Laboratory (LNGS) in Italy. It consists of a cryostat housing the target crystals shielded from the external radioactivity by a water tank acting, at the same time, as an active veto against cosmic ray-induced events.  
    Taking into account both environmental radioactivity and intrinsic contamination of materials used for cryostat, shielding and infrastructure, we performed a careful background budget estimation. The goal is to evaluate the number of events that could mimic or interfere with signal detection while optimising the geometry of the experimental setup.
    In this paper we present the results of the detailed Monte Carlo simulations we performed, together with the final design of the setup that minimises the residual amount of background particles reaching the detector volume.
\end{abstract}

\section{Introduction}

According to the Standard Cosmological Model \linebreak($\Lambda$CDM), Dark Matter (DM) accounts for about $84.4\,$\% of the total matter density of the Universe. The presence of a dark and non-baryonic matter species is confirmed by a variety of cosmological and astronomical observations. However, a definitive proof of the existence of DM particles either in indirect, direct or accelerator searches, is still missing~\cite{PhysRevD.98.030001}.   

Direct searches aim for the detection of DM-induced scattering events in a given target. In this field the DAMA/LIBRA experiment~\cite{Bernabei:2008yh}, operating 250~kg of high purity tallium-doped sodium iodide (NaI(Tl)) crystals at room temperature, reports a statistically robust \linebreak(\unit[12.9]{$\sigma$}~\cite{Bernabei:2018yyw}) annual modulation of the event rate, with period and phase (1 yr and 152.5 d) compatible with a halo of DM particles in the milky way \cite{Lee:2013xxa}. Nevertheless, this result is not confirmed by any other DM experiment. To cross check the DAMA/LIBRA claim in a model-independent way, several experiments using the same target material as DAMA/LIBRA (i.e. NaI crystals) are planned (SABRE~\cite{Antonello_2019}, PICO-LON~\cite{fushimi_dark_2016}, COSINUS~\cite{angloher_cosinus_2016}) or already taking data (COSINE~\cite{Adhikari:2019off}, ANAIS~\cite{Amare:2019jul}). Among them, COSINUS is the only one using a unique detector technology to operate the NaI crystals as scintillating calorimeters (see scheme of detector module in Fig.~\ref{fig:ModuleScheme}): a double cryogenic read-out channel approach provides the possibility to simultaneously detect the scintillation light and the phonon signal, thus allowing to disentangle e$^{-}$/$\gamma$ events from nuclear recoils on an event-by-event basis. This particle identification technique is a crucial benefit, as most DM models predict DM-nucleus scattering, while e$^{-}$/$\gamma$ events are the dominant background.

\section{The COSINUS Experiment}\label{section_COSINUSexp}

\begin{figure}[t]
    \centering
    \includegraphics[width=\linewidth]{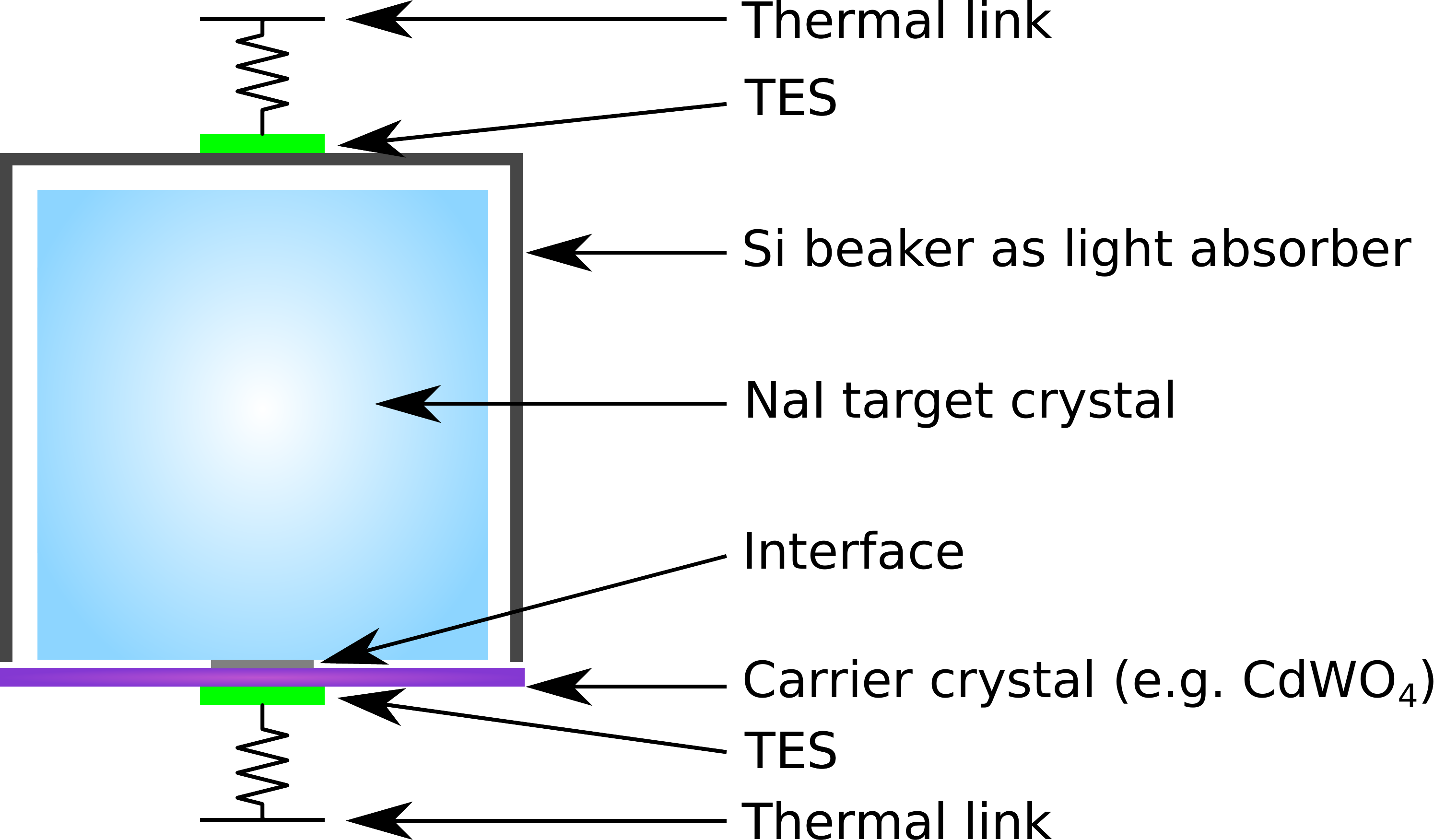}
	\caption{Scheme of a COSINUS detector module consisting of a NaI target crystal coupled to a carrier crystal (phonon detector) and a beaker-shaped light detector. Both channels are read out via a Transition Edge Sensor (TES).}
	\label{fig:ModuleScheme}
\end{figure}

\begin{figure}
    \centering
    \includegraphics[width=\linewidth]{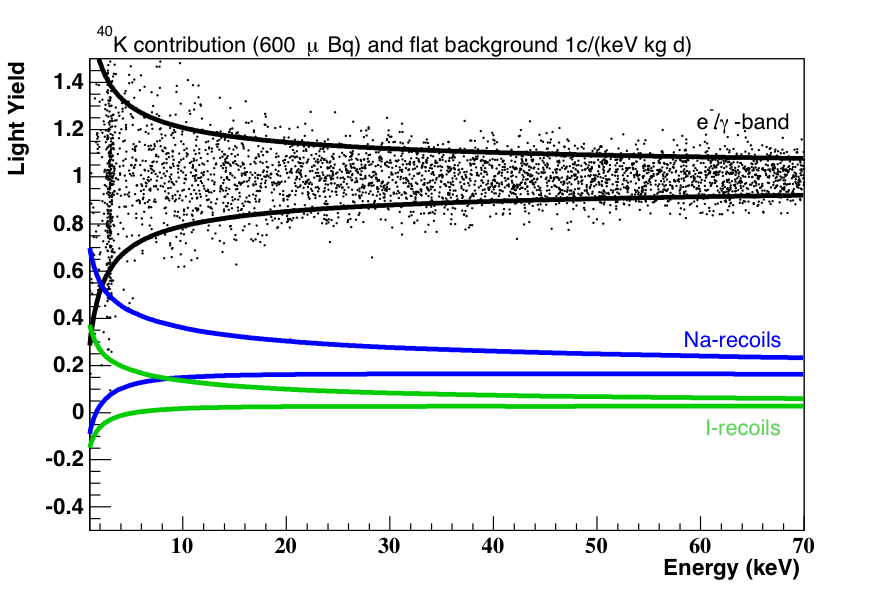}
    \caption{Simulated COSINUS data in the light yield versus energy plane for a detector reaching the performance goals presented in \cite{angloher_cosinus_2016} and assuming a background level compatible with DAMA/LIBRA: a flat electromagnetic background of \unit[1]{\ctskgdkeV} and an internal $^{40}$K contamination of \unit[600]{$\upmu$Bq} responsible for the  X-ray line at $\sim$\unit[3]{keV}. The solid lines mark the upper and lower \unit[90]{\%} boundaries of the e$^{-}$/$\gamma$-band (black) and the nuclear recoil bands for recoils off sodium (blue) and iodine (green), respectively. This plot is based on the energy-dependent quenching factors reported in \cite{tretyak_semi-empirical_2010}. Plot adopted from \cite{angloher_cosinus_2016}.}
    \label{fig:ly_simulation}
\end{figure}

The largest part of the energy deposited by a particle interacting in a scintillating crystal is converted into lattice vibrations (phonons). These are measurable as heat when operating the detector close to absolute zero (T $\sim$ $\mathcal{O}$(mK)). At the same time, a smaller energy fraction of $\mathcal{O}$(\unit[1-10]{\%}) is emitted in the form of scintillation light. While the amount of scintillation light produced by e$^{-}$/$\gamma$ events strongly differs from that produced by nuclear recoils, the phonon (heat) signal is almost independent of the interacting particle type and can be used to precisely measure the deposited energy. We define the Light Yield (LY) as the ratio between light and phonon signal to quantify the separation between e$^{-}/\gamma$ and nuclear recoil events on an event-by-event basis. By definition, we set the LY of e$^{-}$/$\gamma$ particles to 1 at the energy of the calibration source (typically \unit[122]{keV} gammas from Co-57). Nuclear recoils and $\alpha$s feature a lower light yield quantified by particle- and energy-dependent Quenching Factor (QF) values. The QF is defined as the ratio between the scintillation light produced by the energy deposition of a given particle and the amount of light produced by the interaction of e$^{-}$/$\gamma$ particles of the very same energy. Fig.~\ref{fig:ly_simulation} shows the simulated event distribution in the LY versus energy plane, assuming a flat electromagnetic background of \unit[1]{\ctskgdkeV} and an internal $^{40}$K contamination of \unit[600]{$\upmu$Bq}. For further details the reader is referred to \cite{angloher_cosinus_2016}.

 In the baseline design of the COSINUS detector modules (Fig.~\ref{fig:ModuleScheme}), the NaI target is coupled to a highly sensitive temperature sensor via a thin ($\sim$\unit[1]{mm}) carrier crystal made of a different material (e.g.~CdWO$_4$). The interface between carrier and target crystal is made of epoxy resin or silicone oil. The thermometer is a Transition Edge Sensor (TES) consisting of a superconducting tungsten thin film (W-TES, \unit[200]{nm}) produced via electron-beam evaporation. Over the past 20 years, the CRESST dark matter search \cite{CRESST-III_2019_firstResults} has developed and optimised this W-TES technology. The TES is stabilised in its transition between the normal and the superconducting phase. In this configuration the small temperature rise ($\mathcal{O}$($\upmu$K)), due to the energy deposited by a particle interaction in the crystal, is followed by a steep increase in the TES resistance of $\mathcal{O}$(m$\Upomega$), which is in turn registered by Superconducting Quantum Interference Device (SQUID) amplifiers.
 
 A beaker-shaped silicon light absorber of $\sim$\unit[40]{mm} in height and diameter and a wall-thickness of $\sim$\unit[500]{$\upmu$m}, encapsulates the NaI crystal. This configuration enhances the light collection efficiency while acting, at the same time, as an active veto against surface $\alpha$-events. The read-out is performed by a TES optimised for the light detector which is directly evaporated onto the silicon surface. Both light and phonon channel TESs are thermally linked to the heat bath. For a summary of the performance obtained during the detector R$\&$D phase, the reader is referred to~\cite{Angloher:2017sft,Reindl:2017bun,Schaffner:2018jhw,DiMarco:2018gpz,DiMarco:2019oda}.
 
\begin{figure}
    \centering
    \includegraphics[width=0.48\textwidth]{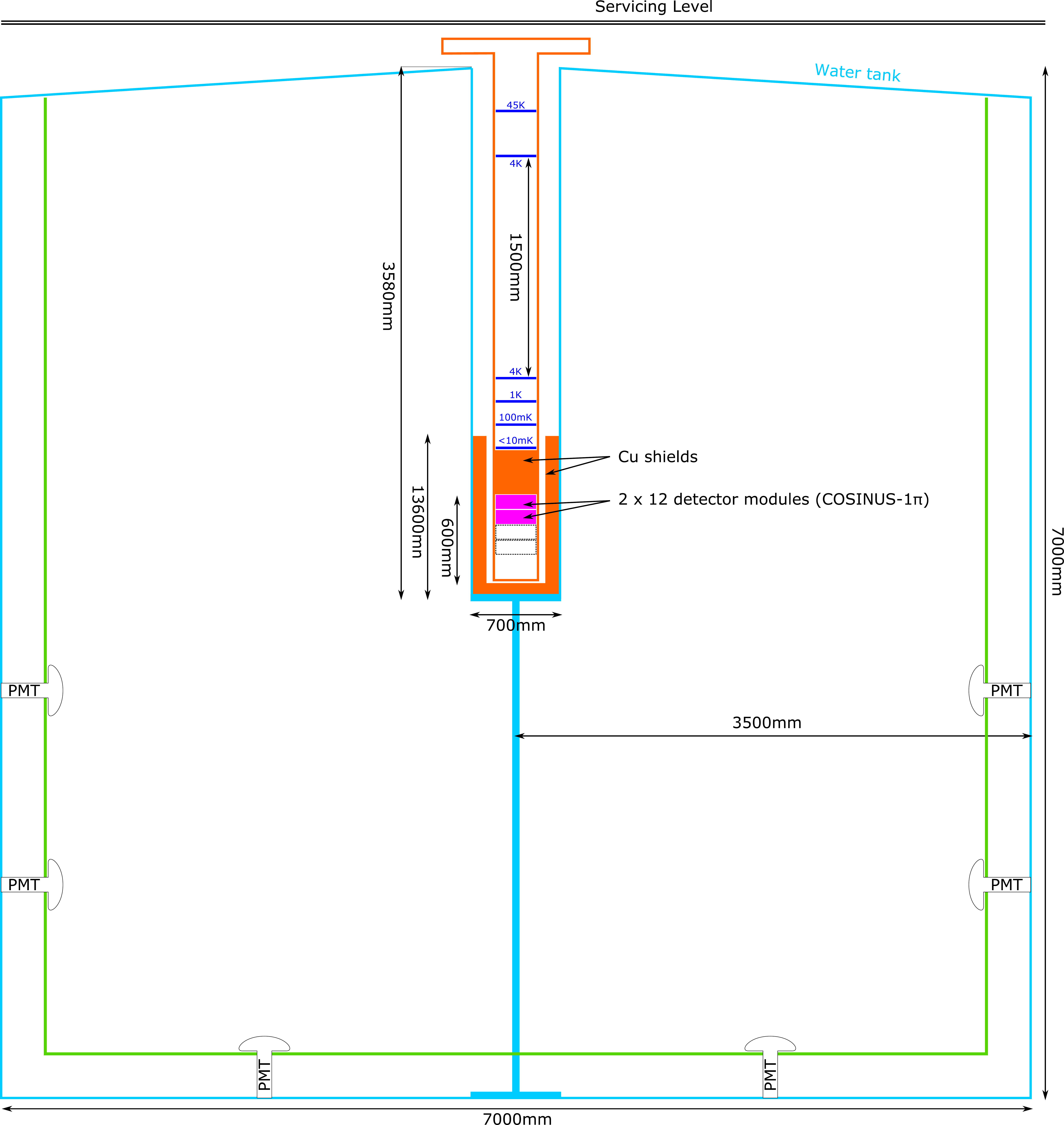}
    \caption{Simplified scheme of the cryostat in the dry-well of the water tank. Light blue: stainless steel walls of water tank and dry-well. Orange: copper parts (shielding and cryostat). Solid magenta, dashed black: experimental volume needed for COSINUS-1\textpi~ and -2\textpi, respectively. Blue: temperature stages of the cryostat. Green: light-tight curtain to create a passive layer at the water tank walls. This scheme is based on the favoured solution (Option 4) for the shielding concept as found by MC simulations (see section \ref{sec:results}).}
    \label{fig:tank_simple}
\end{figure}

The modules described above, composed of both phonon and light detectors, will be arranged in an air-tight copper container to protect the NaI crystals, which are highly hygroscopic, from ambient, humid air. The target volume can host up to six layers of detectors, each housing twelve modules equipped with NaI crystals with a maximum weight of \unit[150]{g} per module. A simplified scheme of the overall experimental apparatus is shown in Fig.~\ref{fig:tank_simple}. The space for the target is allocated at the bottom of a cylindrical stainless steel tube, the ``dry-well''. This tube is inserted in a tank filled with ultra-pure water and hosts the cryostat. The geometry is conceived to place the detector volume at the center of the tank. For this reason, the dry dilution refrigerator is equipped with a custom-made extension. The water tank will be equipped with photomultiplier tubes (PMTs), thus acting at the same time as passive shielding against ambient radiation and  active Cherenkov veto against cosmogenic background. The experimental apparatus will be installed in Hall B of the LNGS underground laboratory.

The COSINUS experiment is planning a phased approach. In the first phase, called COSINUS-1$\pi$, up to two layers of detectors (i.e. 24 modules) will be installed. The goal is to provide, exploiting the particle identification capability of our detectors, a cross-check of the DAMA/LIBRA result independent of the dark matter halo (astrophysics) and of the interaction mechanism (particle physics) apart from demanding that dark matter interacts with nuclei and not with electrons in the target. As discussed in detail in~\cite{Kahlhoefer:2018knc}, assuming a nuclear recoil threshold on the phonon signal of about \unit[1]{keV} and the radiopurity level and performance discussed in~\cite{angloher_cosinus_2016}, the exposure needed to fulfil the COSINUS-1$\pi$ objective is of the order of \unit[1000]{\kgd}. Depending on the final mass of the single NaI crystal, not yet fixed, the goal can be achieved running about 20 crystals, \unit[50(100)]{g} each, for \unit[3(1.5)]{years}.

In the second phase, called COSINUS-2$\pi$, the search for the annual modulation signal is planned by increasing the target mass up to the maximum capacity (72 modules in total). 

The reach of the physics goal of the project imposes the need of an accurate background budget estimation taking into account both environmental and intrinsic radiation. The objective of the setup is to efficiently shield the former with appropriate materials, while reducing the amount of the latter originating from the unavoidable radioactive contamination contained in both shielding and infrastructure. 

\section{Background sources and shielding concept} \label{sec:bkg_budget}

\begin{table*} [t]
 	\centering
 	\caption{The five different options for our shielding configuration, featuring different thicknesses of water, Pb, Cu and PE. See Fig.~\ref{fig:ShieldOptions} for a schematic view of the examined configurations.}
 \begin{tabular}{c c c c c c c c c c c} 
 \hline
 Option    & Tank                  & Water 	    	& Dry-well          & \multicolumn{3}{c}{Inner shielding} & Cryostat &  \multicolumn{3}{c}{Top shielding}   \\ [0.5ex] 
    & (steel)           & radius            & (stainless steel)       &    (Pb) & (Cu) & (PE)   & (Cu)     & (Pb) & (Cu) & (PE)     \\
        & [cm]                  &  [cm]             & [cm]              & [cm] & [cm] & [cm] & [cm] & [cm] &  [cm] &  [cm] \\ 
 \hline
  1       & 1.5 & 150 & 0.4 & 10 & 15 & 10 & 0.8 & 10 & 15 & 10 \\ 
  2       & 1.5 & 200 & 0.4 &  0 & 15 & 10 & 0.8 & 0 & 40 & 10 \\
  3       & 1.5 & 200 & 0.4 &  0 & 15 &  0 & 0.8 & 0 & 40 & 0 \\
  4       & 1.5 & 300 & 0.4 &  0 &  8 &  0 & 0.8 & 0 & 30 & 0 \\
  5       & 1.5 & 300 & 0.4 &  0 &  0 &  0 & 0.8 & 0 & 40 & 0 \\
 \hline
\end{tabular}
\label{tab:DetailDesignOptions}
\end{table*}

	\begin{figure*} [t]
			\centering
			\begin{subfigure}[b]{0.6\linewidth}
				\includegraphics[width=\linewidth]{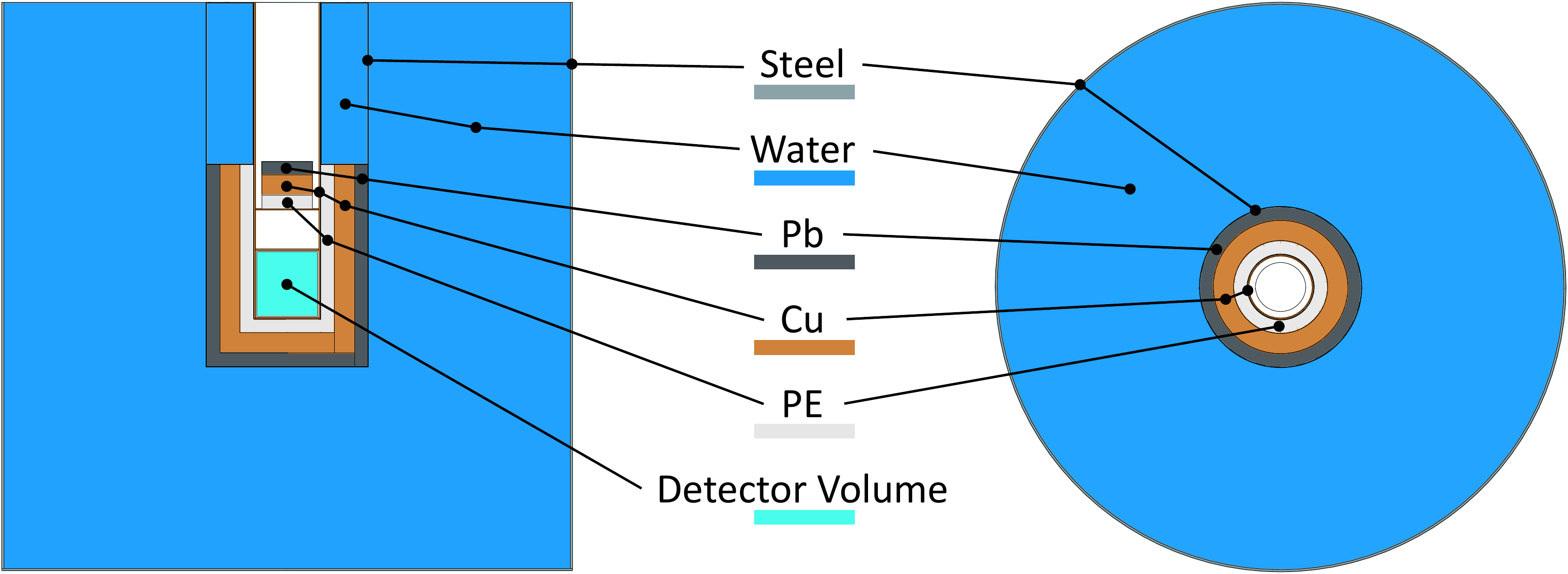}
				\caption{Option 1. Left: lateral view. Right: top view}
            \label{fig:setup-a}
			\end{subfigure}
			
			\vspace{4px}
			\begin{subfigure}[b]{0.22\linewidth}
				\includegraphics[width=\linewidth]{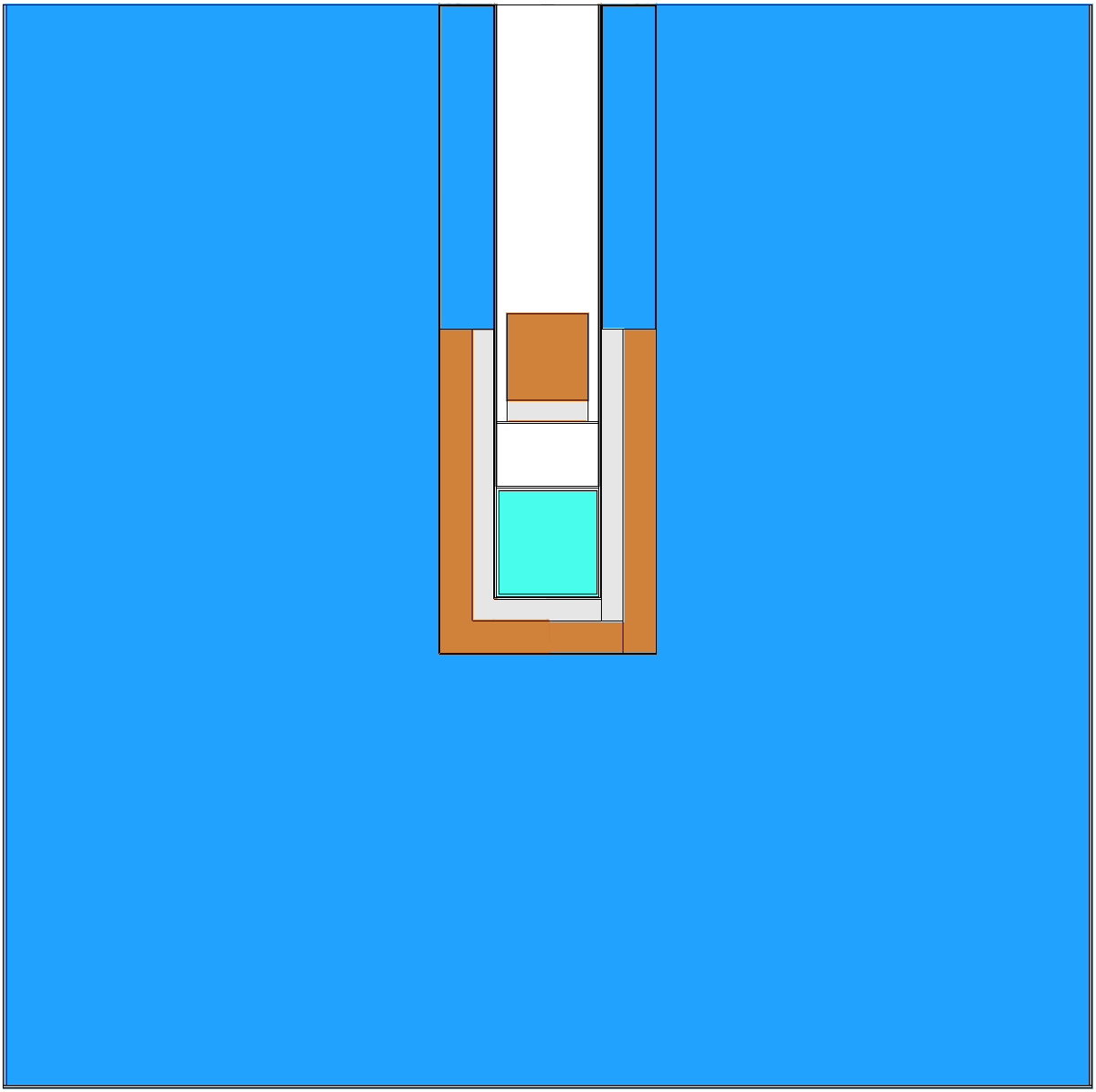}
				\caption{Option 2.}
				\label{fig:setup-b}
			\end{subfigure}
			\begin{subfigure}[b]{0.22\linewidth}
				\includegraphics[width=\linewidth]{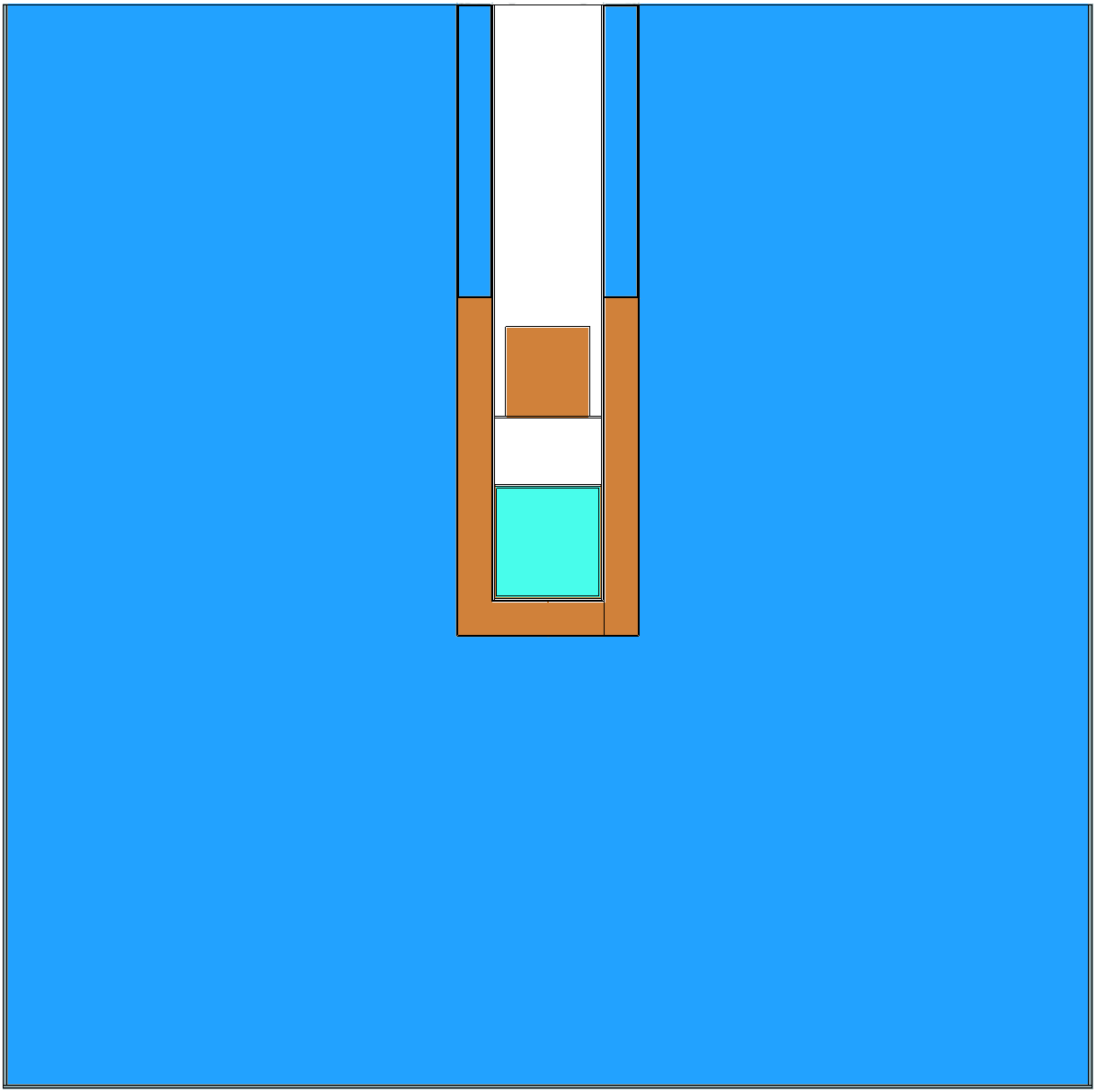}
				\caption{Option 3.}
				\label{fig:setup-c}
			\end{subfigure}
			\begin{subfigure}[b]{0.22\linewidth}
				\includegraphics[width=\linewidth]{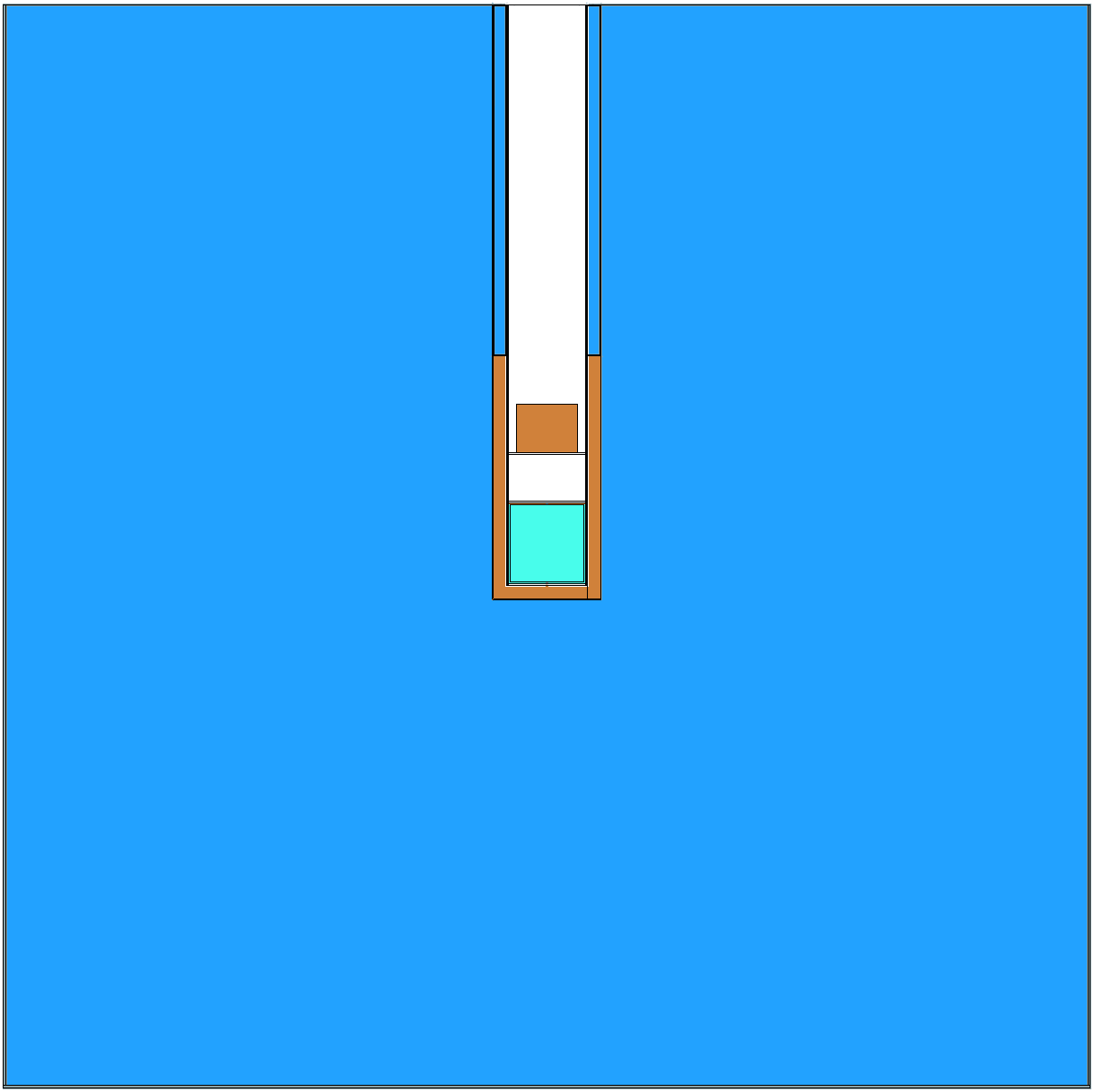}
				\caption{Option 4.}
				\label{fig:setup-d}
			\end{subfigure}
			\begin{subfigure}[b]{0.22\linewidth}
				\includegraphics[width=\linewidth]{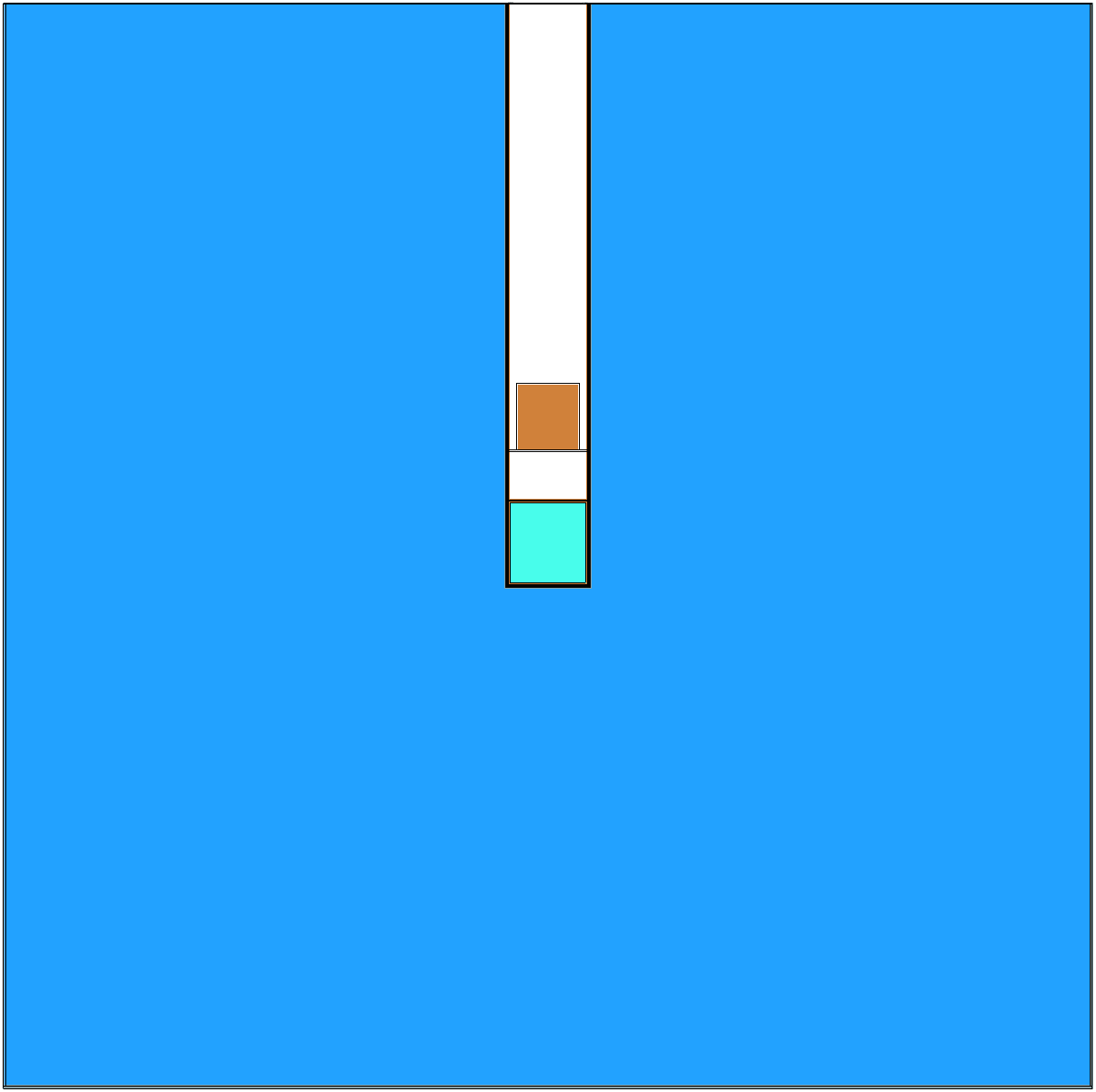}
				\caption{Option 5.}
				\label{fig:setup-e}
			\end{subfigure}
			\caption{The five different experimental setups considered for the MC simulation. The Cu cryostat containing the detector volume (cyan box), which will house the NaI detectors, is inserted in a thin stainless steel structure (dry-well) hosting the shielding layers. The detector volume is positioned at the center of a water-filled stainless steel tank. Details about the thickness of the various layers for the five setup options can be found in Tab.~\ref{tab:DetailDesignOptions}.}
			\label{fig:ShieldOptions}
		\end{figure*}    

COSINUS is the unique NaI experiment featuring a discrimination capability between e$^{-}/\gamma$-particles and nuclear recoil events on an event-by-event basis. In the search for DM particles scattering off atomic nuclei, neutrons interacting in the target crystals can mimic a DM signal and thus have to be considered as the most dangerous type of background. Nevertheless, $\gamma$ radioactivity has to be properly taken into account especially in the very low energy region where the discrimination power is weaker (see Fig.~\ref{fig:ly_simulation}). 

Neutrons, in underground sites, can be produced in reactions initiated both by natural radioactivity (radiogenic neutrons) and cosmic rays (cosmogenic neutrons). Radiogenic neutrons are produced via  spontaneous fission and $(\alpha,n)$ interactions originating from natural radioactive contaminants (i.e. $^{238}$U and $^{232}$Th) in rock, water and concrete surrounding the apparatus as well as in the materials used in the experimental setup. Cosmogenic neutrons are instead produced either in muon-induced spallation processes or by secondary particles generated in muon-induced cascades, i.e.~through photon- or hadron-induced spallation or disintegration reactions~\cite{Kudryavtsev:2008fi}. While the radiogenic neutron energy is in the fast range (up to $\sim$\unit[10]{MeV}), cosmogenic neutrons have energies extending up to few GeV.  

Ambient $\gamma$s originate from long-lived natural radioisotopes $^{40}$K, $^{238}$U and $^{232}$Th present in the rock. Moreover, cosmic muons can induce $\gamma$ radiation via bremsstrahlung both in the rock and in high-Z materials possibly used as passive shielding. 

The general strategy in reducing the flux of ambient and radiogenic neutrons and $\gamma$s is based on the use of successive layers of passive absorbers. An outermost layer of a low-A material (typically polyethylene (PE) or water) acts as an efficient moderator for ambient neutrons. To reduce the environmental $\gamma$ flux, an appropriate thickness of a high-Z material (usually lead) is needed. A layer of copper is then used to mitigate the $\gamma$ radiation emitted in the decay of the long-lived $^{210}$Pb isotope contained in lead. Finally, to deal with radiogenic neutrons produced in the shielding materials, a further layer of a low-A material is often used as the innermost shield close to the detector volume. 
In addition to these layers, the natural shielding provided by the rock overburden above the LNGS underground halls (3600 m.w.e.~\cite{ambrosio_vertical_1995}), where the COSINUS apparatus will be located, assures a reduction of the muon flux by a factor of $\sim$10$^6$ with respect to the surface \cite{LVD98}. Nonetheless, especially in high-Z materials like Pb used as $\gamma$ absorbers, muon-induced spallation and subsequent hadronic showers result in a further emission of neutrons. A muon veto to tag such events is therefore crucial. The use of a tank filled with ultra-pure water and instrumented with PMTs satisfies both, the requirement for an external passive shield absorbing ambient radiation and the need for an active veto. Moreover, ultra-pure water has a lower contamination level than PE and a proper tuning of the water thickness allows reducing the amount of high-Z material layers, thus finally minimising the sources of radiogenic as well as those of muon-induced backgrounds. 

\begin{figure*} [t]
			\centering
			\begin{subfigure}[b]{0.45\linewidth}
				\includegraphics[width=\linewidth]{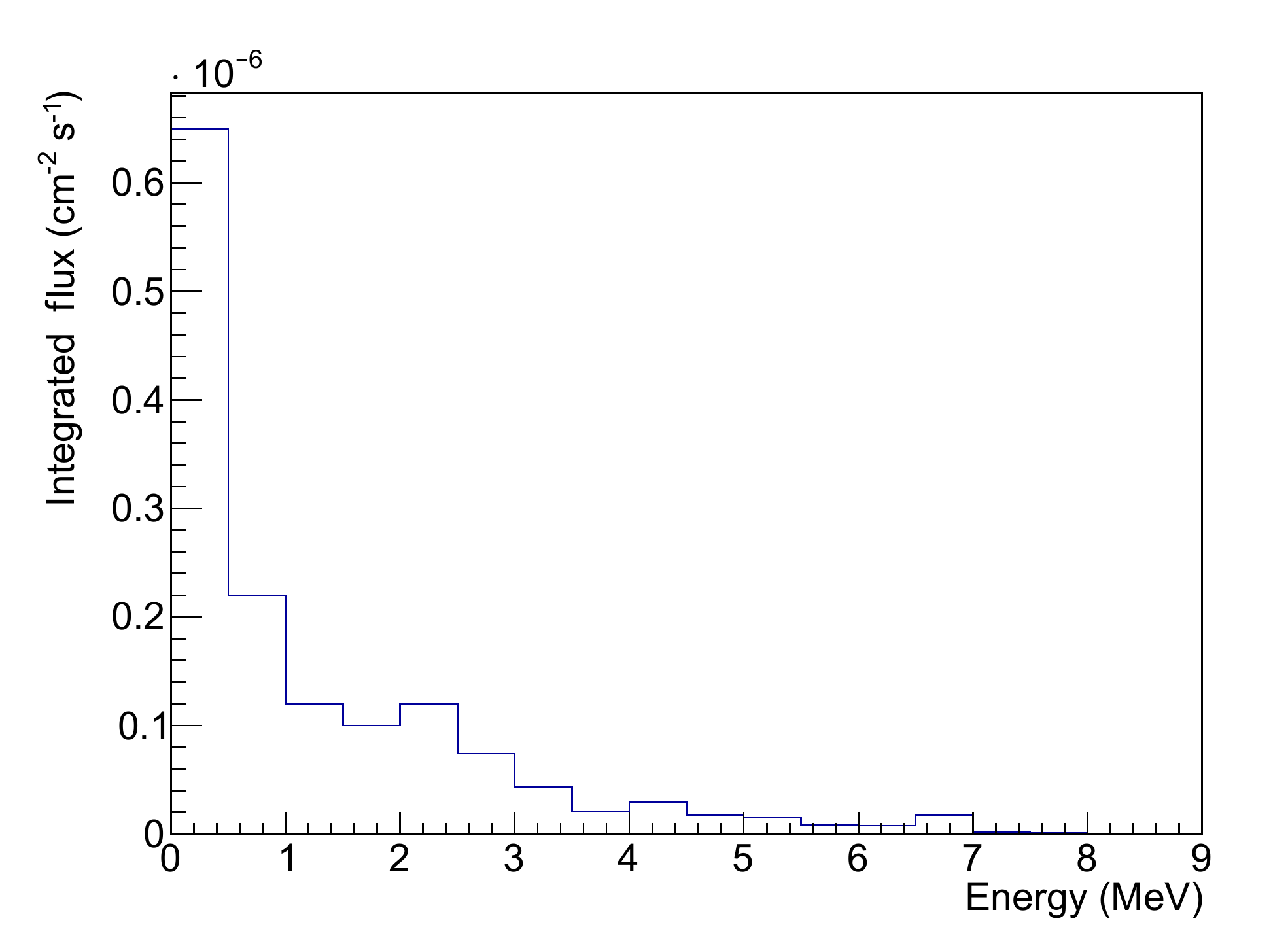}
				\caption{Ambient neutron flux according to~\cite{Wulandari:2003cr}.}
            \label{fig:AmbientNeutronSpectrum}
			\end{subfigure}
			\begin{subfigure}[b]{0.45\linewidth}
				\includegraphics[width=\linewidth]{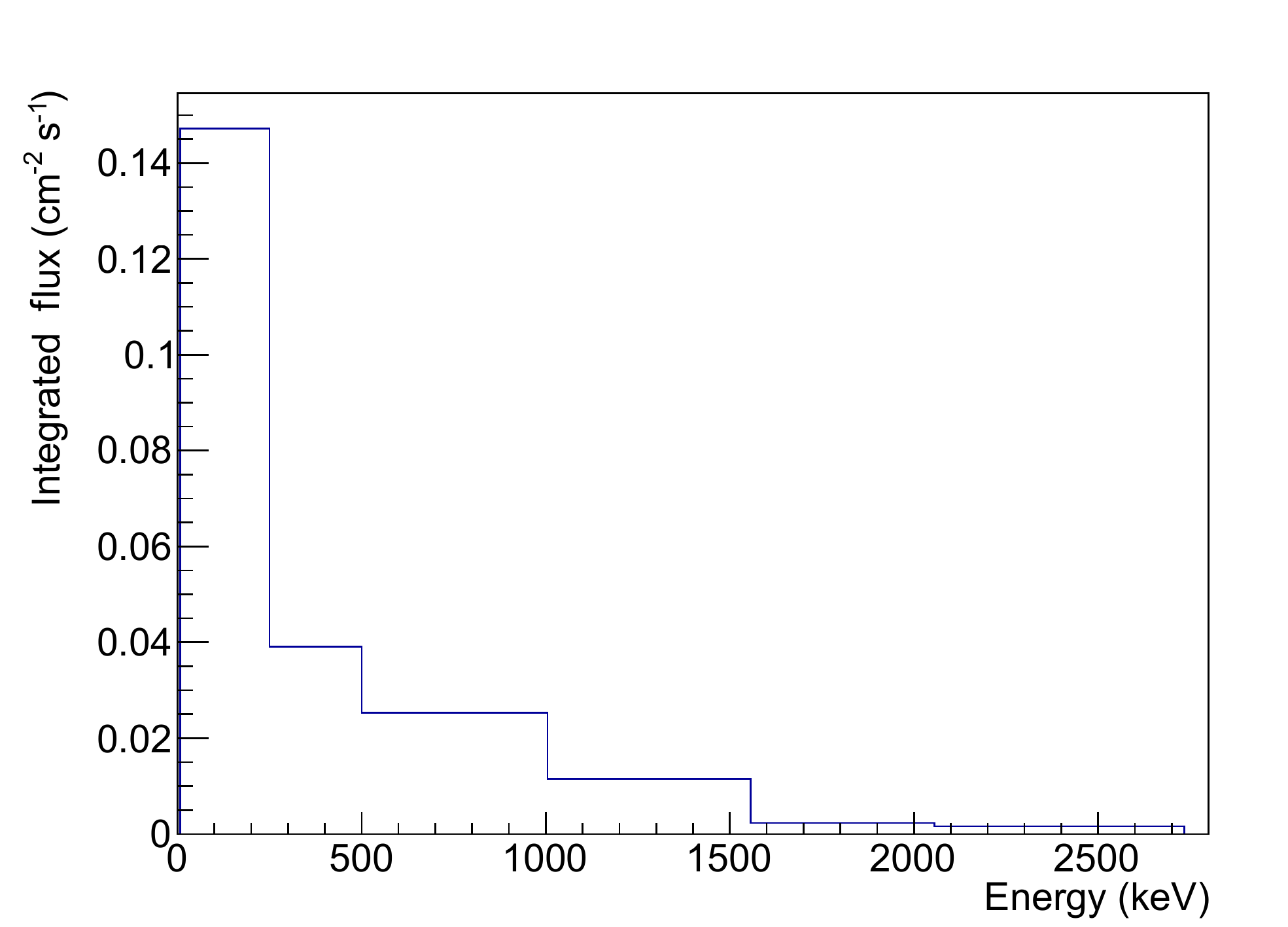}
				\caption{Ambient gamma flux (simplified) according to \cite{MalAmbientGammaLNGS}.}
				\label{fig:AmbientGammaSpectrum}
			\end{subfigure}
			\begin{subfigure}[b]{0.45\linewidth}
				\includegraphics[width=\linewidth]{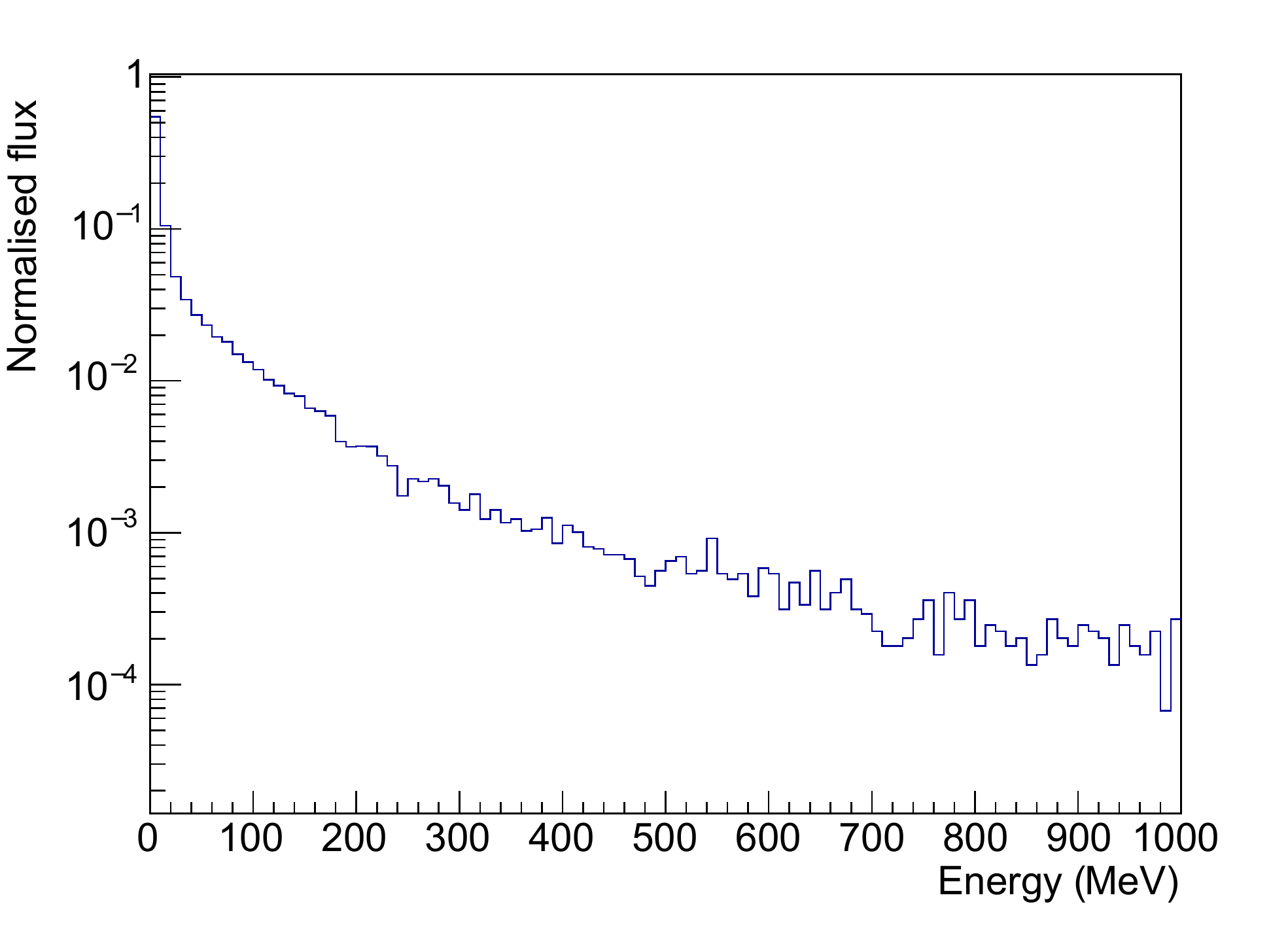}
				\caption{Muon-induced neutron spectrum.}
				\label{fig:MuIndNeutronSpectrum}
			\end{subfigure}
			\begin{subfigure}[b]{0.45\linewidth}
				\includegraphics[width=\linewidth]{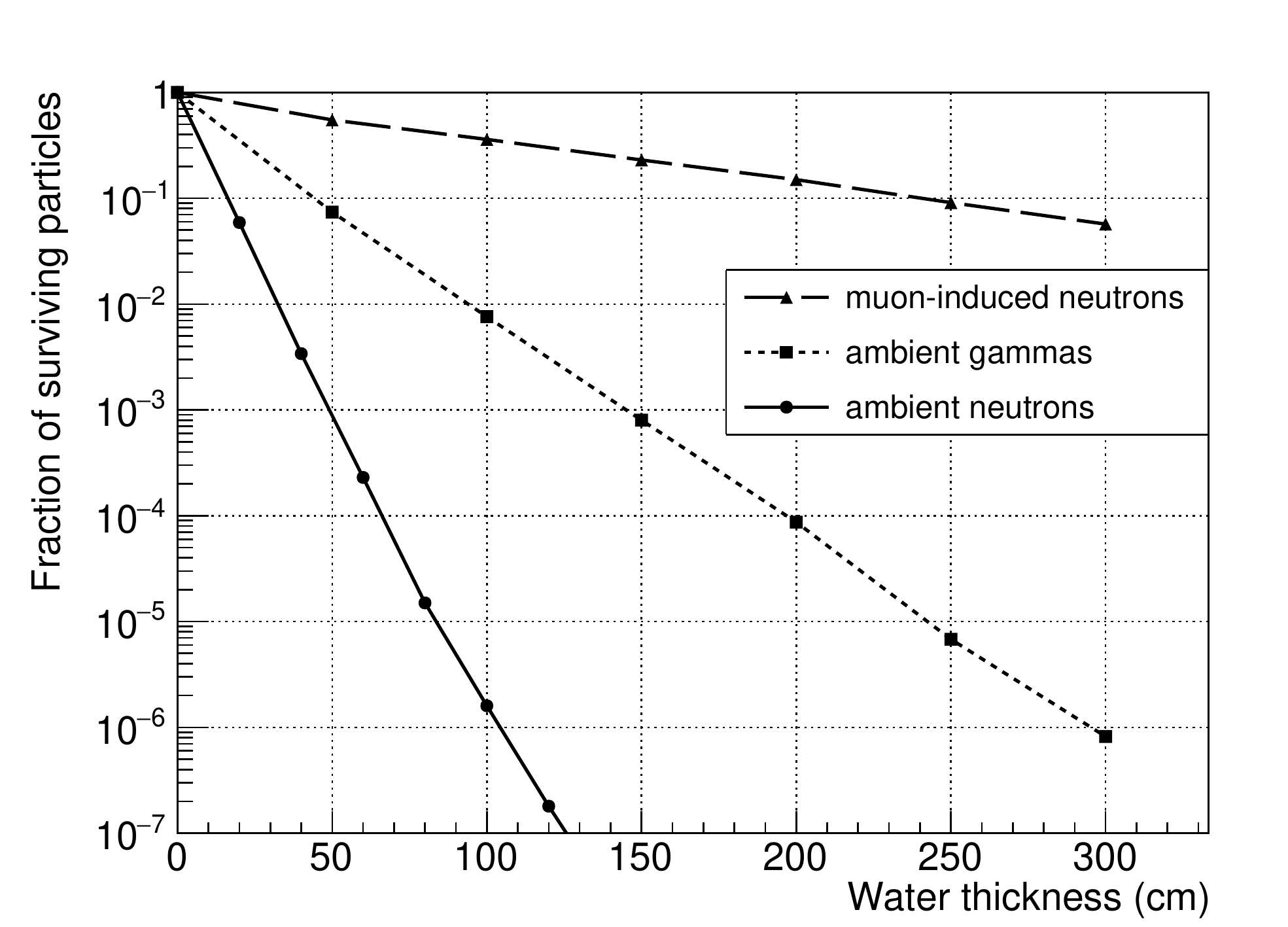}
				\caption{Particles transmitted through water.}
				\label{fig:SurvivingParticlesThroughWater}
			\end{subfigure}
			\caption{The plots show spectra of background components external to our experimental setup and the respective fractions of surviving particles as a function of the thickness of the water layer. The rather low-energy ambient neutrons are much more efficiently shielded than comparably high-energy muon-induced neutrons. The efficiency for shielding ambient $\gamma$s with water lies between the other two values.}
			\label{fig:ExternalBackgroundMitigation}
		\end{figure*}

\section{Simulation}\label{section_Simulation}
To estimate the overall number of particles reaching the detector volume in our setup, we performed Monte Carlo (MC) simulations based on Geant4 (v10.2.3)~\cite{AGOSTINELLI2003250,Allison:2006ve,ALLISON2016186}, SOURCES4C~\cite{SOURCES} and MUSUN~\cite{Kudryavtsev:2008qh}.

\begin{table*} [t]
 	\centering
 	\caption{Activity of the main contaminants in the materials considered for the COSINUS setup. Secular equilibrium is assumed for the decay chains. The rightmost column states the neutron yield due to $(\alpha,n)$ and spontaneous fission reactions attained using the SOURCES4C code~\cite{SOURCES}.}
 	\label{tab:MaterialActivities}
 \begin{tabular}{c c c c c c c c c c } 
 \hline
 Material & $^{238}$U & $^{235}$U & $^{232}$Th & Ref. & $^{40}$K & $^{60}$Co & $^{137}$Cs    & Ref.                & Neutron yield				\\ [0.5ex] 
          & [mBq/kg]  & [mBq/kg]  & [mBq/kg] & & [mBq/kg]  & [mBq/kg]  & [mBq/kg]  &
          & [cm$^{-3}$ s$^{-1}$] 	\\ 
 \hline
 Stainless steel    & $<0.2$     & --\textsuperscript{*}        & $<0.1$ & \cite{Artusa:2014wnl}   & $<5.2$ & $1.9$ & $<0.6$       &  \cite{Aprile:2017ilq}  & $3.0 \cdot 10^{-12}$ \\ 
 Pb       & $<0.01$     & --\textsuperscript{*}        & $<0.07$ & \cite{Artusa:2014wnl}  & -- & -- & --      &  --  &  $1.2 \cdot 10^{-13}$ 									\\
 Cu       & $<0.065$    & --\textsuperscript{*}        & $<0.002$ & \cite{Alduino:2016vjd}  & $<0.34$ & $0.21$ & $<0.03$     &  \cite{Aprile:2017ilq} & $6.6 \cdot 10^{-13}$ \\
 PE       & $< 3.8$     & $<0.37$   & $<0.14$ & \cite{Aprile_2011}  & $0.7$ & $<0.1$ & $0.06$      &  \cite{Aprile_2011}  & $9.4 \cdot 10^{-12}$ \\ [1ex]
 \hline
 \multicolumn{6}{l}{\textsuperscript{*}no measured value given, natural abundance assumed\hfill}
\end{tabular}
\end{table*}

Starting from the basic configuration sketched in Fig.~\ref{fig:tank_simple} and detailed in section~\ref{section_COSINUSexp}, we simulated five simplified experimental layouts as shown in Fig.~\ref{fig:ShieldOptions}. The cryostat, represented by a single Cu layer with a thickness of \unit[0.8]{cm}, containing the detector volume (cyan box), is inserted in a thin (\unit[0.4]{cm}) steel structure, the dry-well, which also hosts the inner passive shielding layers. They may consist of one or two layers of high-Z materials (Pb and Cu, or Cu only) and potentially a further layer of polyethylene (PE). The five configurations depicted in Fig.~\ref{fig:ShieldOptions} differ for the thickness of shielding materials used as listed in detail in Tab.~\ref{tab:DetailDesignOptions}. 

For each material and its relative thickness used in the different design options shown in Fig.~\ref{fig:ShieldOptions}, we compute the radiogenic neutron and $\gamma$ fluxes, assuming the intrinsic radioactive contamination values quoted in literature, as reported in Tab.~\ref{tab:MaterialActivities}. The intrinsic neutron and $\gamma$ yields are used as a benchmark and compared to the attenuation of the environmental fluxes offered by the various shielding configurations to optimise the thicknesses of the different layers. The goal of the optimisation is to reduce the ambient background to a negligible level with respect to the unavoidable intrinsic radiogenic background. 

The background rates will nonetheless depend on the final detector layout (e.g. size and number of modules) which is currently under definition. A detailed background model will therefore be the topic of future work. For the preferred shielding option, however, we will briefly discuss the effect of detector size and segmentation at the end of this section.
    
Besides using passive shields, an active muon tagging system will be crucial to reduce the dominating neutron background contribution presented in section \ref{section_CosmogenicNeutrons}. For this purpose, we will use the water tank as an active Cherenkov veto. Its tagging efficiency will depend on the size of the tank and on the number, distribution, trigger condition and performance of PMTs used to measure the Cherenkov light.

\subsection{Ambient neutrons}\label{section_AmbientNeutrons}

The energy spectrum of ambient neutrons measured in the LNGS underground halls, taken from Ref.~\cite{Wulandari:2003cr}, is shown in Fig.~\ref{fig:AmbientNeutronSpectrum} in bins of 500$\;$keV. The integral flux in the range 1$-$500$\,$keV is \unit[$6.5 \cdot 10^{-7}$]{\fluxcms}, while the integrated flux above 500$\;$keV is \unit[$7.9 \cdot 10^{-7}$]{\fluxcms}.

We performed a Geant4-based simulation, propagating neutrons, generated according to the above spectrum, through a large water cuboid. We conservatively considered neutrons entering perpendicularly to the water surface. The fraction of surviving ambient neutrons as a function of the thickness of the water layer is reported in Fig.~\ref{fig:SurvivingParticlesThroughWater}. An attenuation factor of $10^{5}$ ($10^{7}$) can be obtained for a thickness of less than \unit[1]{m} (\unit[1.5]{m}). The flux of ambient neutrons reaching the detector volume, in any of the considered shielding configurations, will thus be lower than \unit[$10^{-13}$]{\fluxcms}. Projecting this number to the surface of the detector region would translate to less than \unit[$3.5 \cdot 10^{-2}$]{yr$^{-1}$} entering the detector volume. The ambient neutron flux is hence reduced to a negligible level with respect to the potential background sources presented in the subsequent sections. Moreover, surviving neutrons are efficiently thermalised, so that they do not contribute to the background via nuclear recoils.
        
To cope with particles impinging the target volume perpendicularly from the top, through the neck of the dry-well where no water is present, we foresee to deploy a layer of a low-Z material around the extension of the cryostat (\unit[1.5]{m}) between the first two cooling stages and the dilution unit (see Fig.~\ref{fig:tank_simple}). For details on the cryostat design the reader is referred to~\cite{COSINUS_CDR}.
This layer will either be composed of PE or could consist of thin PE tanks filled with ultra-pure water and shaped to fit the space around the cryostat prolongation. The latter solution would assure a lower intrinsic radioactive contamination level and facilitate the mechanical handling when lifting the cryostat out of the dry-well.

\begin{figure} [t]
			\centering
			\includegraphics[width=\linewidth]{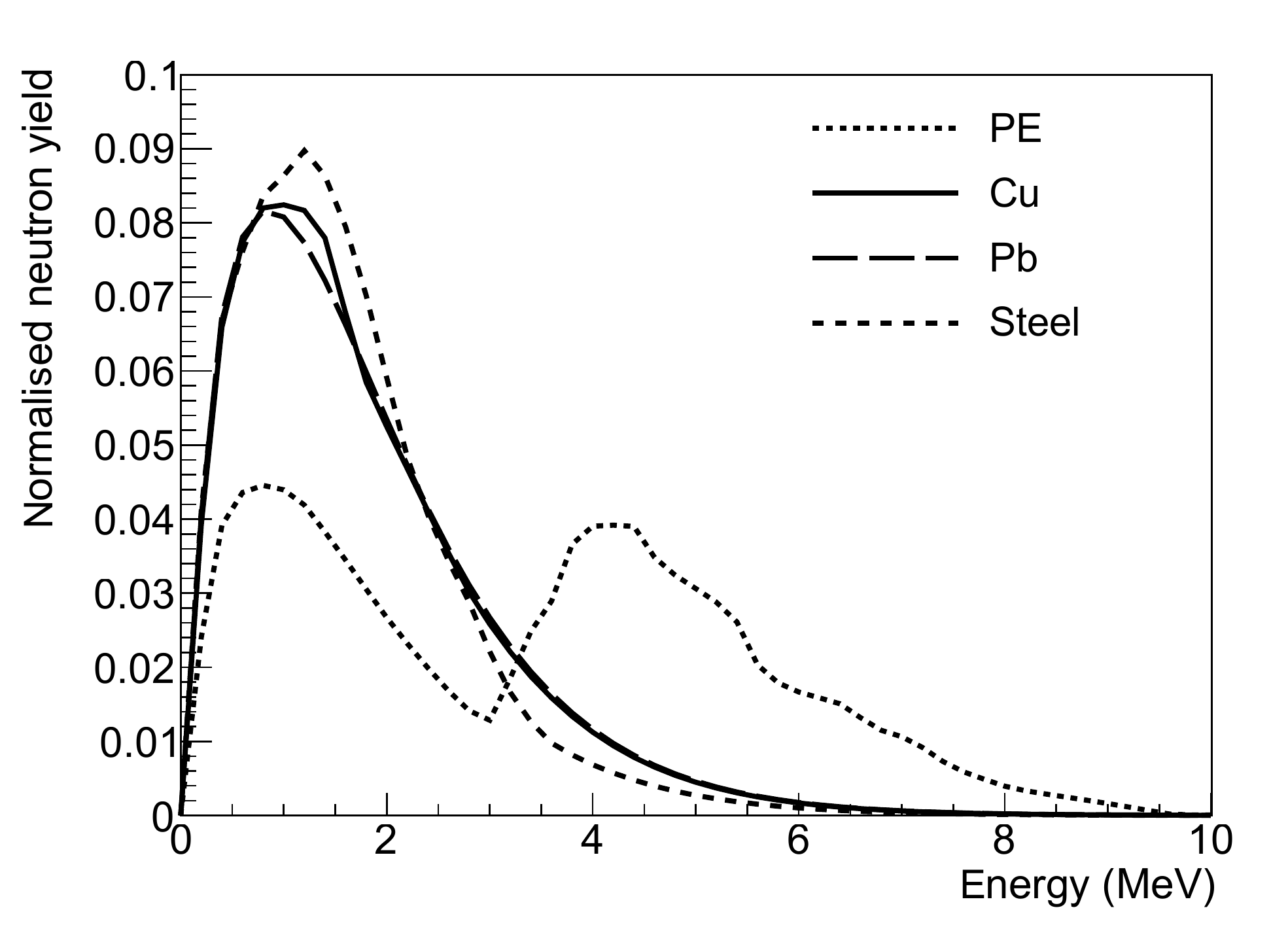}
			\caption{Energy spectrum of radiogenic neutrons obtained using the SOURCES4C code~\cite{SOURCES} for PE, Pb and Cu shielding layers. The double peak in the energy distribution of PE is due to the presence of $^{13}$C, which has a high cross section for $(\alpha,n)$ interactions. All spectra are normalized to unit integral.}
			\label{fig:RadiogenicNeutronSpectrum}
\end{figure}        
        
\subsection{Radiogenic neutrons}\label{section_RadiogenicNeutrons}
	
For the intrinsic activity of the materials employed in the various shielding layers and in the infrastructure of the COSINUS apparatus, we assumed the upper limits reported by other rare event searches listed in Tab.~\ref{tab:MaterialActivities}. To estimate the flux and the energy spectrum of neutrons originating in the various materials via spontaneous fission and ($\alpha$,n) reactions, we performed a simulation using the SOURCES4C code \cite{SOURCES} and assuming secular equilibrium for $^{238}$U and $^{232}$Th decay chains. The integrated radiogenic neutron yield is reported, for each material, in the last column of Tab.~\ref{tab:MaterialActivities}, while the energy spectra are shown in Fig.~\ref{fig:RadiogenicNeutronSpectrum}. These results are used as input for the Geant4 simulation. 

The result of the simulation is reported in Tab.~\ref{tab:RadiogenicNeutrons_Results}: for each shielding material within a given design option, we evaluate the number of neutrons reaching the target volume. The sum of all contributions is reported in the last column of the table. 
        
As expected, less employed material leads to a lower number of radiogenic neutrons reaching the detector volume. Moreover, PE, despite its small mass fraction, results in the largest contribution to the total radiogenic neutron flux, $\mathcal{O}$(\unit[$10$]{events/yr}), due to the higher contamination levels with respect to other shielding materials (see Tab.~\ref{tab:MaterialActivities}) and higher ($\alpha$,n) yield.
        
Radiogenic neutrons originating from the intrinsic radioactivity of target crystals do not contribute to the overall background. Neutrons produced in ($\alpha$,n) reactions inside the crystal can indeed be tagged by detecting the preceding $\alpha$ particle having a typical energy in the order of a few MeV, thus inducing a saturated signal. A clearly saturated signal is also obtained for intrinsic neutrons derived from spontaneous fission, as they are emitted together with nuclear fragments with energies of the order of \unit[$\sim$100]{MeV}. 
        
\subsection{Cosmogenic neutrons}\label{section_CosmogenicNeutrons}
\begin{figure} [t]
	\centering
		\begin{subfigure}[b]{0.95\linewidth}
			\includegraphics[width=\linewidth]{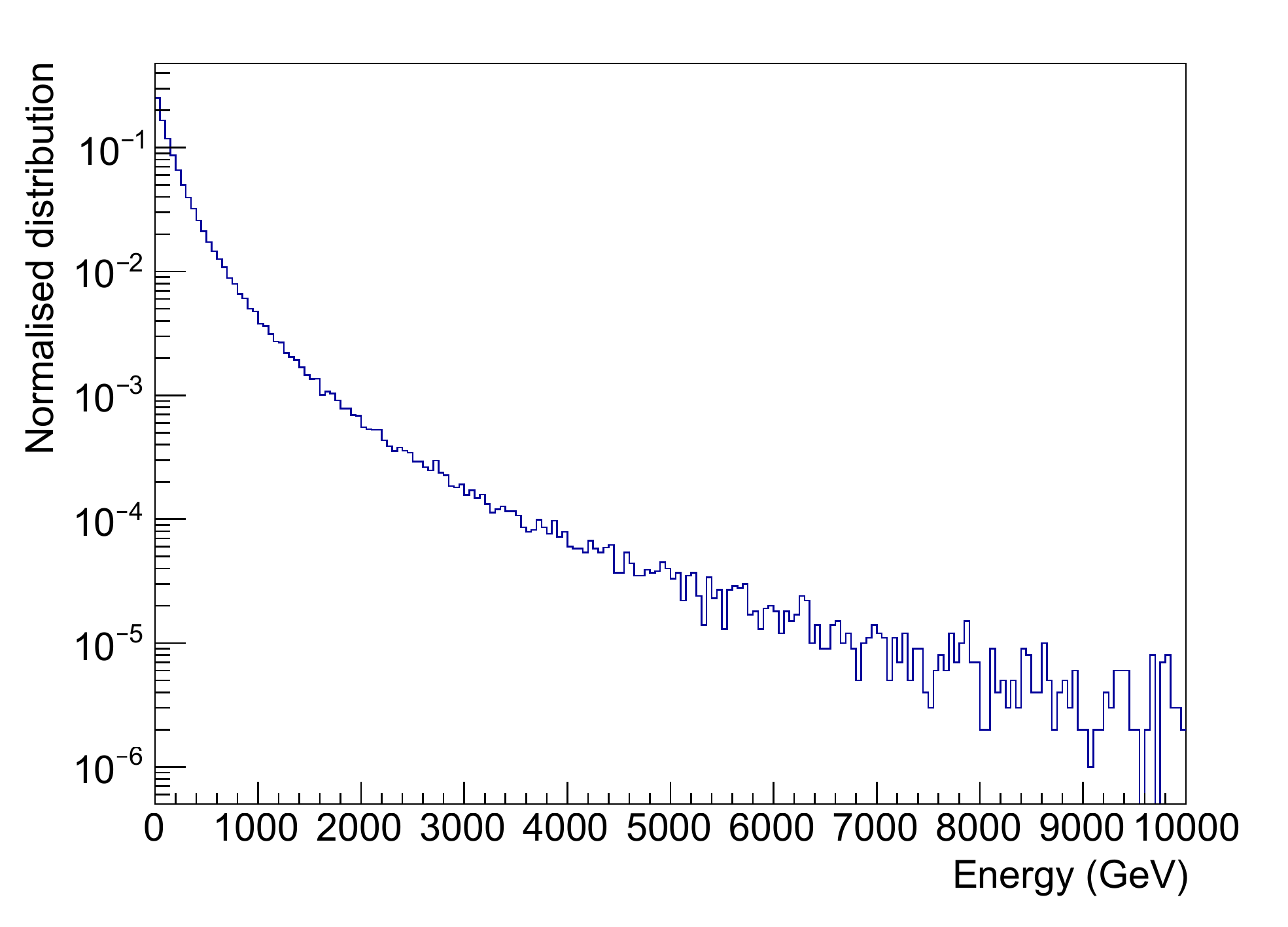}
			\vspace{-20px}
			\caption{}
			\vspace{5px}
		\end{subfigure}
		\begin{subfigure}[b]{0.45\linewidth}
			\includegraphics[width=\linewidth]{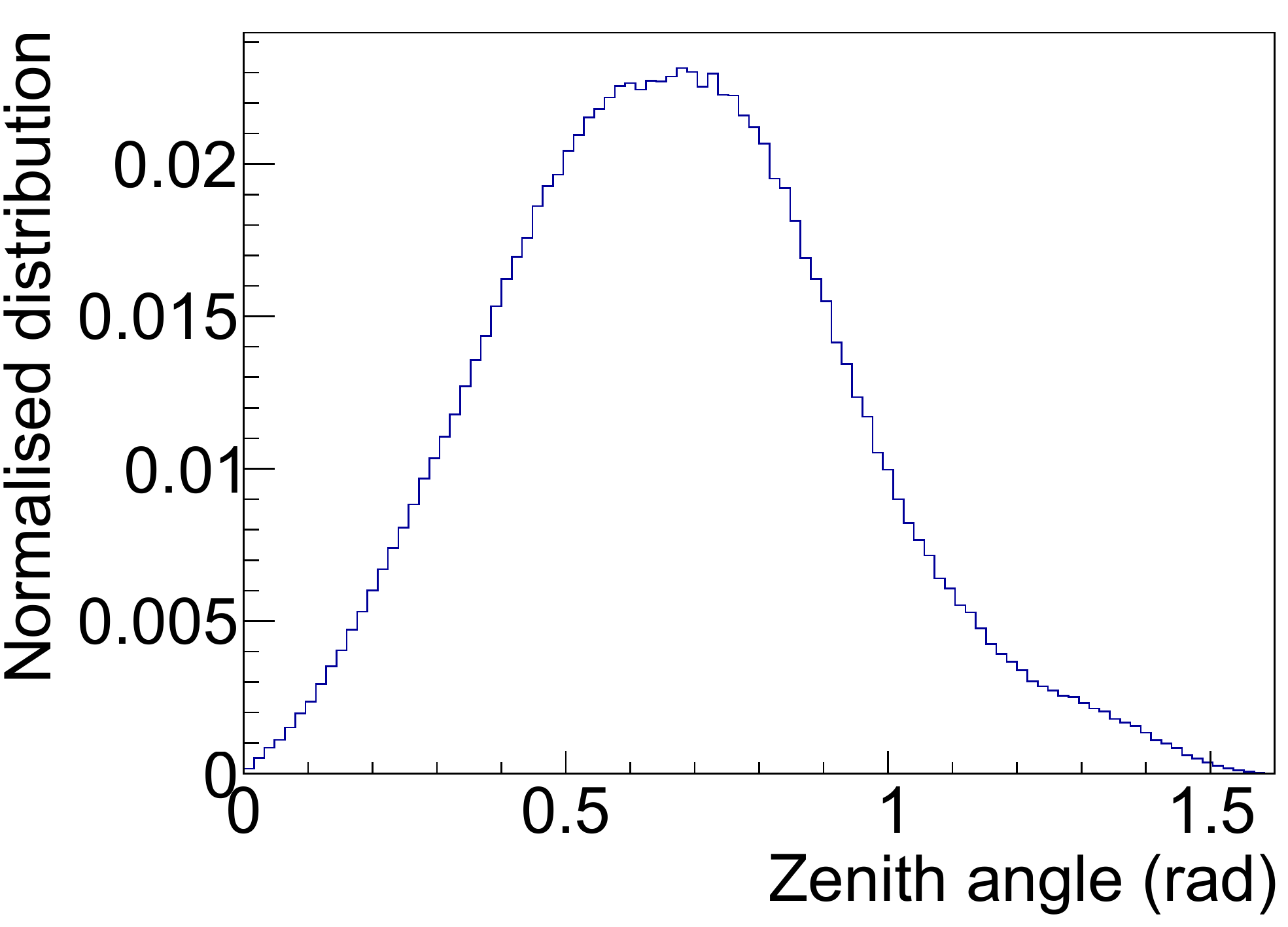}
			\vspace{-15px}
			\caption{}
		\end{subfigure}
		\begin{subfigure}[b]{0.45\linewidth}
			\includegraphics[width=\linewidth]{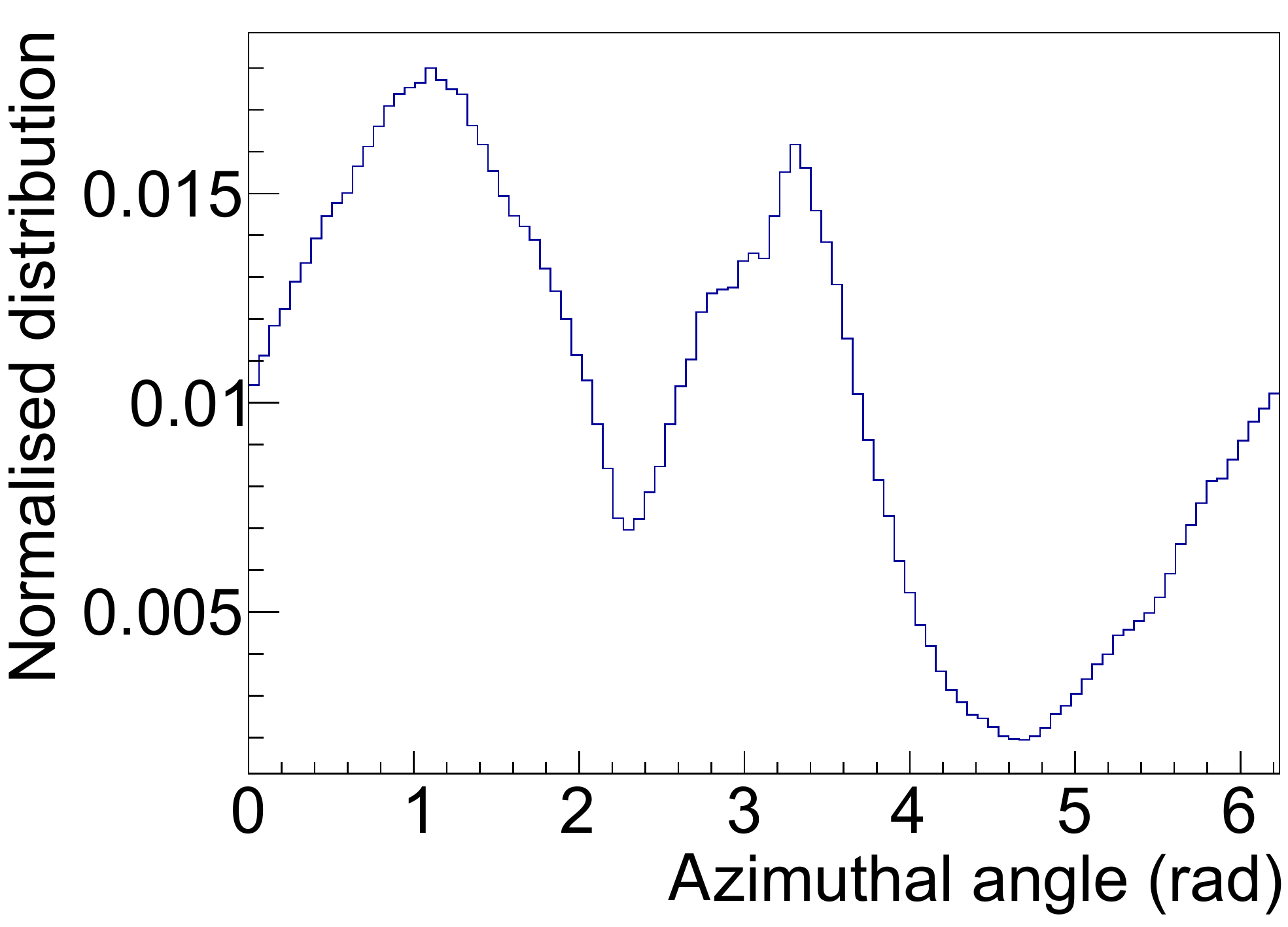}
			\vspace{-15px}
			\caption{}
		\end{subfigure}
			\caption{Spectrum of the (a) energy, (b) zenith angle and (c) azimuthal angle of muons reaching the LNGS underground laboratory, obtained using the MUSUN \cite{Kudryavtsev:2008qh} simulation code.}
		\label{fig:MuonSpectr_MUSUN}
\end{figure}

As mentioned in section~\ref{sec:bkg_budget}, muons reaching the underground halls at LNGS are suppressed by a factor of\linebreak$\sim10^6$ with respect to sea level thanks to the \unit[3600]{m.w.e.} shielding provided by the rock overburden \cite{ambrosio_vertical_1995,LVD98}. Nevertheless, surviving muons produce neutrons through spallation processes in the rock or in high-Z materials surrounding the detector. To estimate the cosmogenic contribution to the overall background, we performed a simulation of the cosmic muon flux reaching the underground site by using the MUSUN code~\cite{Kudryavtsev:2008qh}. The results obtained are compatible with values found in literature~\cite{Kudryavtsev:2008fi}: the mean energy of muons reaching the underground halls is \unit[270]{GeV}, the average zenith angle is \unit[0.67]{rad} and both the distribution of the zenith and azimuthal angles are compatible with the profile of the Gran Sasso mountain (see Fig.~\ref{fig:MuonSpectr_MUSUN}). 

\begin{table*}
	\centering
	\caption{Radiogenic neutron background contributions originating from distinct materials used in the various design options. Stated values include 1 $\sigma$ statistical uncertainties.}
	\label{tab:RadiogenicNeutrons_Results} 
	\begin{tabular}{||c c c c ||} 
		\hline
		\multirow{2}{*}{\rule{0pt}{0ex}Design option}   & \multirow{2}{*}{\rule{0pt}{0ex}Neutron source} & Events with neutrons    & \multirow{2}{*}{\rule{0pt}{0ex}Total}             											\\ 
		& & entering detector volume & \\[0.5ex]
		&				& [yr$^{-1}$] & [yr$^{-1}$] 	\\ 
		\hline\hline
		\multirow{4}{*}{\rule{0pt}{4.5ex}Option 1}& PE      	& $(8.86 \pm 0.01) \cdot 10^{0}$ 	& \multirow{4}{*}{\rule{0pt}{4.5ex}$(9.17 \pm 0.01) \cdot 10^{0}$}	\\ [0.2ex]
		\cline{2-3}\rule{0pt}{2.5ex}
		& Cu      			& $(2.96 \pm 0.03) \cdot 10^{-1}$ &																				\\ [0.2ex]
		\cline{2-3}\rule{0pt}{2.5ex}
		& Pb      			& $(7.41 \pm 0.46) \cdot 10^{-3}$  	&																				\\ [0.2ex]
		\cline{2-3}\rule{0pt}{2.5ex}
		&Steel    			& $(3.62 \pm 0.34) \cdot 10^{-3}$ &																				\\
		\hline
		\multirow{3}{*}{\rule{0pt}{3.75ex}Option 2}  & PE      			& $(8.91 \pm 0.01) \cdot 10^{0}$  & \multirow{3}{*}{\rule{0pt}{3.75ex}$(9.18 \pm 0.01) \cdot 10^{0}$}	\\ [0.2ex]
		\cline{2-3}\rule{0pt}{2.5ex}
		& Cu      			& $(2.67 \pm 0.03) \cdot 10^{-1}$  &																				\\ [0.2ex]
		\cline{2-3}\rule{0pt}{2.5ex}
		&Steel    			& $(2.06 \pm 0.24) \cdot 10^{-3}$ &																				\\ 
		\hline
		\multirow{2}{*}{\rule{0pt}{2.5ex}Option 3}  & Cu      			& $(2.16 \pm 0.01) \cdot 10^{0}$  & \multirow{2}{*}{\rule{0pt}{2.5ex}$(2.17 \pm 0.01) \cdot 10^{0}$}	\\ [0.2ex]
		\cline{2-3}\rule{0pt}{2.5ex}
		&Steel    			& $(1.19 \pm 0.06) \cdot 10^{-2}$ &																				\\ 
		\hline
		\multirow{2}{*}{\rule{0pt}{2.5ex}Option 4}  & Cu      			& $(8.44 \pm 0.05) \cdot 10^{-1}$ & \multirow{2}{*}{\rule{0pt}{2.5ex}$(9.31 \pm 0.07) \cdot 10^{-1}$}	\\ [0.2ex]
		\cline{2-3}\rule{0pt}{2.5ex}
		&Steel    			& $(8.72 \pm 0.16) \cdot 10^{-2}$  &																		\\
		\hline
		\multirow{2}{*}{\rule{0pt}{2.5ex}Option 5} & Cu (cryostat) & $(2.09 \pm 0.02) \cdot 10^{-1}$ & \multirow{2}{*}{\rule{0pt}{2.5ex}$(4.22 \pm 0.05) \cdot 10^{-1}$}	\\ [0.2ex]
		\cline{2-3}\rule{0pt}{2.5ex}
		&Steel     			& $(2.13 \pm 0.03) \cdot 10^{-1}$  &																				\\ 
		\hline
	\end{tabular}
\end{table*}

\begin{table*}
	\centering
	\caption{Cosmogenic neutron background contributions in the various design options. Stated values include 1 $\sigma$ statistical uncertainties.}
	\label{tab:CosmogenicNeutrons_Results} 
	\begin{tabular}{|| c c ||} 
		\hline
		\multirow{2}{*}{\rule{0pt}{0ex}Design option}   & Events with neutrons  \\
		& entering detector volume \\ [0.5ex]
		& [yr$^{-1}$]  									
		\\ 
		\hline\hline
		Option 1 & $(2.10 \pm 0.03) \cdot 10^{2}$  \\
		\hline 
		Option 2 & $(1.15 \pm 0.02) \cdot 10^{2}$  \\  
		\hline
		Option 3 & $(3.36 \pm 0.04) \cdot 10^{2}$  \\
		\hline
		Option 4 & $(2.22 \pm 0.03) \cdot 10^{2}$  \\
		\hline
		Option 5 & $(1.11 \pm 0.02) \cdot 10^{2}$  \\
		\hline
	\end{tabular}
\end{table*}
	
As output of the MUSUN simulation we obtained an integral flux of \unit[$2.3\cdot 10^6$]{yr$^{-1}$}, through the surface of a cuboid with dimension $12\times 12\times 13\;$m$^3$: position, direction and energy of each simulated muon are stored and used as input for the Geant4 simulation. Muons are first propagated through a thickness of at least \unit[2.5]{m} of rock (to properly take into account the interaction with the rock surrounding the apparatus), and then through the five different options of our experimental setup listed in Tab.~\ref{tab:DetailDesignOptions}. We assumed the water tank positioned at a fixed distance of \unit[0.5]{m} from one of the walls of the underground hall.

\begin{table*}
	\centering
	\caption{Radiogenic $\gamma$s, originating from distinct materials used in the various design options, entering the detector volume. Stated values include 1 $\sigma$ statistical uncertainties.}
	\label{tab:RadiogenicGammas_Results} 
	\begin{tabular}{||c c c c ||} 
		\hline
		\multirow{2}{*}{\rule{0pt}{0ex}Design option}   & \multirow{2}{*}{\rule{0pt}{0ex}Gamma source} & Events with gammas    & \multirow{2}{*}{\rule{0pt}{0ex}Total}             											\\ 
		& & entering detector volume & \\[0.5ex]
		&				& [yr$^{-1}$] & [yr$^{-1}$] 	\\ 
		\hline\hline
		\multirow{4}{*}{\rule{0pt}{4.5ex}Option 1}& PE    	& $(5.41 \pm 0.13) \cdot 10^{6}$  	& \multirow{4}{*}{\rule{0pt}{4.5ex}$(5.68 \pm 0.14) \cdot 10^{6}$}	\\ [0.2ex]
		\cline{2-3}\rule{0pt}{2.5ex}
		& Cu     			& $(2.74 \pm 0.11) \cdot 10^{5}$ &																				\\ [0.2ex]
		\cline{2-3}\rule{0pt}{2.5ex}
		& Pb      			& $(8.90 \pm 0.89) \cdot 10^{2}$  	&																				\\ [0.2ex]
		\cline{2-3}\rule{0pt}{2.5ex}
		& Steel    			& $(2.70 \pm 0.69) \cdot 10^{1}$ &																				\\
		\hline
		\multirow{3}{*}{\rule{0pt}{3.75ex}Option 2}  & PE  	& $(5.41 \pm 0.13) \cdot 10^{6}$  & \multirow{3}{*}{\rule{0pt}{3.75ex}$(5.68 \pm 0.14) \cdot 10^{6}$}	\\ [0.2ex]
		\cline{2-3}\rule{0pt}{2.5ex}
		& Cu	  			& $(2.74 \pm 0.11) \cdot 10^{5}$ &																				\\ [0.2ex]
		\cline{2-3}\rule{0pt}{2.5ex}
		& Steel    			& $(1.16 \pm 0.10) \cdot 10^{3}$ &																				\\ 
		\hline
		\multirow{2}{*}{\rule{0pt}{3.ex}Option 3}  & Cu  	& $(4.06 \pm 0.12) \cdot 10^{5}$  & \multirow{2}{*}{\rule{0pt}{3.ex}$(4.08 \pm 0.13) \cdot 10^{5}$}	\\ [0.2ex]
		\cline{2-3}\rule{0pt}{2.5ex}
		&Steel    			& $(2.36 \pm 0.15) \cdot 10^{3}$ &																		\\
		\hline
		\multirow{2}{*}{\rule{0pt}{3.ex}Option 4}  & Cu  	& $(3.97 \pm 0.11) \cdot 10^{5}$  & \multirow{2}{*}{\rule{0pt}{3.ex}$(4.46 \pm 0.14) \cdot 10^{5}$}	\\ [0.2ex]
		\cline{2-3}\rule{0pt}{2.5ex}
		&Steel    			& $(4.89 \pm 0.25) \cdot 10^{4}$ &																		\\
		\hline
		\multirow{2}{*}{\rule{0pt}{3.ex}Option 5} & Cu (cryostat) & $(1.51 \pm 0.02) \cdot 10^{5}$ & \multirow{2}{*}{\rule{0pt}{3.ex}$(2.09 \pm 0.04) \cdot 10^{6}$}	\\ [0.2ex]
		\cline{2-3}\rule{0pt}{2.5ex}
		&Steel     			& $(1.94 \pm 0.04) \cdot 10^{6}$  &																				\\ 
		\hline
	\end{tabular}
\end{table*}
		
For each design option in Fig.~\ref{fig:ShieldOptions}, we simulated between 15 and 30 million muons, roughly corresponding to an exposure between 6.5 and 13 years. Fig.~\ref{fig:MuIndNeutronSpectrum} shows the energy distribution of cosmogenic neutrons produced. As expected, the spectrum extends to much higher values with respect to the radiogenic one (see Fig.~\ref{fig:RadiogenicNeutronSpectrum} for comparison). The number of events in which cosmogenic neutrons enter the detector volume is reported in Tab.~\ref{tab:CosmogenicNeutrons_Results}. The reason for quoting the number of events rather than the number of neutrons is that in a single muon-induced event many neutrons, as part of a hadronic shower, may enter the target volume at the same time. The various design options exhibit comparable performance, since a notable fraction of neutrons is produced in the rock surrounding the setup. Nonetheless, it is worth noting that the use of a Pb layer in the shielding results in a higher amount of muon-induced neutrons, as the neutron yield scales with the atomic mass number as A$^{0.8}$ \cite{Kudryavtsev:2008fi}, while a PE layer provides an additional shielding against these neutrons. Compared to the radiogenic background, however, the cosmogenic contribution is one or two orders of magnitude higher for all the experimental options considered. To not limit our sensitivity we thus plan the use of an active Cherenkov muon veto by instrumenting the water tank with PMTs.

We started dedicated simulations to evaluate the necessary number of PMTs and their arrangement in the tank. The final results will be the subject of a future work. Preliminary analyses already show that a veto system equipped with about 30 PMTs reduces the cosmogenic neutron background by roughly two orders of magnitude, thus reaching the level of the estimated radiogenic one.

\subsection{Radiogenic gammas}\label{section_RadiogenicGammas}
The radiogenic $\gamma$ flux originates from the natural decay chains as well as from additional contaminants like $^{40}$K, $^{60}$Co or $^{137}$Cs. To evaluate this contribution we performed a dedicated Geant4 simulation using the contamination levels stated in Tab.~\ref{tab:MaterialActivities}. The estimated radiogenic $\gamma$ rates are reported in Tab.~\ref{tab:RadiogenicGammas_Results}. The numbers show that PE and steel layers lead to high contributions, if no additional shield is placed closer to the cryostat. Different contaminants may play important roles in different materials. While the largest fraction of events which have its source in PE or Cu originate from the $^{238}$U decay chain, the dominating contribution coming from steel originates from its $^{60}$Co contamination. 
    
\subsection{Ambient gammas}\label{section_AmbientGammas}
For the simulation of the ambient $\gamma$ radioactivity in the LNGS underground halls, we adopted the spectrum shown in Fig.~\ref{fig:AmbientGammaSpectrum} taken from~\cite{MalAmbientGammaLNGS}: the integrated $\gamma$ flux is $\sim$ \unit[0.23]{cm$^{-2}$\,s$^{-1}$}. The gamma spectrum has been used as input for the Geant4 simulation computing the number of $\gamma$s transmitted through the water tank. The attenuation factor as a function of water thickness is reported in Fig.~\ref{fig:SurvivingParticlesThroughWater}: an attenuation factor of the order of $\sim 10^{4}$ ($\sim 10^{6}$) is obtained for \unit[200]{cm} (\unit[300]{cm}) of water thickness. The use of a high-Z material (Pb and/or Cu) can further reduce the residual flux. The resulting particle rates reaching the detector volume are reported in Tab.~\ref{tab:AllBackgroundContributions}. In all shielding options, the thicknesses of the various layers enable a reduction of the ambient $\gamma$ flux down to, or below the level of the estimated radiogenic $\gamma$ background (see section \ref{section_RadiogenicGammas}).        

\begin{table*}
	\centering
	\caption{List of all the background contributions in every shielding option considered in the Geant4 simulations. The numbers are attained using the simplified geometrical setup detailed in section \ref{sec:bkg_budget} and represent the number of events per year, in which at least one particle of the corresponding background source enters the detector volume.  Stated values include 1 $\sigma$ statistical uncertainties.}
	\label{tab:AllBackgroundContributions}
	\resizebox{\textwidth}{!}{%
	\begin{tabular}{c c r@{}l r@{}l r@{}l r@{}l r@{}l } 
		\hline
		\multicolumn{2}{c}{\multirow{2}{*}{Background source}} & \multicolumn{10}{c}{Estimated number of particles entering the detector volume (yr$^{-1}$)} \\ [0.5ex]
		& & \multicolumn{2}{c}{Option 1} & \multicolumn{2}{c}{Option 2} & \multicolumn{2}{c}{Option 3} & \multicolumn{2}{c}{Option 4} & \multicolumn{2}{c}{Option 5} \\ [0.5ex] 
		\hline \\ [-1.4ex]
		\multirow{3}{*}{Neutrons $\begin{dcases} \\ \\ \\ \end{dcases}$} &  Ambient     & $<3.50$ &$\,\cdot\, 10^{-2}$ & $<3.50$ &$\,\cdot\, 10^{-2}$ & $<3.50$ &$\,\cdot\, 10^{-2}$ & $<3.50$ &$\,\cdot\, 10^{-2}$ & $<3.50$ &$\,\cdot\, 10^{-2}$  \\ [0.5ex]
		&  Radiogenic  & $(9.17 \pm 0.01)$ &$\,\cdot\, 10^{0}$ & $(9.18 \pm 0.01)$ &$\,\cdot\, 10^{0}$ & $(2.17 \pm 0.01)$ &$\,\cdot\, 10^{0}$ & $(9.31 \pm 0.07)$ &$\,\cdot\, 10^{-1}$ & $(4.22 \pm 0.05)$ &$\,\cdot\, 10^{-1}$  \\ [0.5ex]
		&  Cosmogenic  & $(2.10 \pm 0.03)$ &$\,\cdot\, 10^{2}$ & $(1.15 \pm 0.02)$ &$\,\cdot\, 10^{2}$ & $(3.36 \pm 0.04)$ &$\,\cdot\, 10^{2}$ & $(2.22 \pm 0.03)$ &$\,\cdot\, 10^{2}$ & $(1.11 \pm 0.02)$ &$\,\cdot\, 10^{2}$  \\ [1.5ex]
		\multirow{2}{*}{Gammas $\begin{dcases} \\ \\ \end{dcases}$} &  Ambient     & $(3.15 \pm 1.41)$ &$\,\cdot\, 10^{3}$ & $(6.81 \pm 1.15)$ &$\,\cdot\, 10^{4}$ & $(7.88 \pm 1.05)$ &$\,\cdot\, 10^{4}$ & $(1.71 \pm 0.57)$ &$\,\cdot\, 10^{4}$ & $(4.94 \pm 0.47)$ &$\,\cdot\, 10^{5}$  \\ [0.5ex]
		&  Radiogenic  & $(5.68 \pm 0.14)$ &$\,\cdot\, 10^{6}$ & $(5.68 \pm 0.14)$ &$\,\cdot\, 10^{6}$ & $(4.08 \pm 0.13)$ &$\,\cdot\, 10^{5}$ & $(4.46 \pm 0.14)$ &$\,\cdot\, 10^{5}$ & $(2.09 \pm 0.04)$ &$\,\cdot\, 10^{6}$  \\ 
		[1ex]
		\hline
	\end{tabular}
	}
\end{table*}

\subsection{Systematic uncertainties}
Besides statistical uncertainties in the Monte Carlo evaluations and the uncertainties related to the assumed literature values of contamination levels, we provide, in the following, an estimation of systematic uncertainties in the physics models of employed simulation codes.

SOURCES4C carries an uncertainty in the amount of radiogenic neutrons produced via ($\alpha$,n) reactions. While spontaneous fission is well-described, ($\alpha$,n) cross-sections are based on tabulated databases, which may comprise differing values. In \cite{Mendoza2020}, various codes for calculating $\alpha$-induced neutron yields are compared and SOURCES4C is found to agree to experimental data within \unit[20]{\%} in most of the cases.

MUSUN was tested against data of various experiments and found to be consistent with their results \cite{Kudryavtsev:2008qh}. Particularly, the code was validated against measured data of the LVD experiment \cite{Aglietta1998}, which is situated in hall A of LNGS, thus yielding a negligible systematic uncertainty in the computed amount, energy spectrum and angular distribution of muons reaching the LNGS underground halls.

Uncertainties in Geant4 mainly persist in the muon-induced neutron yield and in data-driven physics models. In \cite{Reichhart2013}, measurement and simulation of the muon-induced neutron production in a deep underground laboratory are checked against each other. The setting is comparable to the LNGS, so that the quoted agreement within \unit[25]{\%} may be assumed as an estimate on the systematic error. For the neutron transport below \unit[20]{MeV}, we employ the data-driven high precision neutron transport model. A comparison in \cite{Lemrani2006} between Geant4 and MCNPX \cite{MCNPX} showed an agreement better than \unit[20]{\%}, which we can assume to be an upper limit on the systematic error on low-energy neutron propagation.
    
\section{Discussion of the results}\label{sec:results}

In Tab.~\ref{tab:AllBackgroundContributions}, the rate of particles reaching the detector volume in the considered shielding options is listed. Looking at these results and considering the discussions in the previous sections, we can draw the following conclusions.
    
Neutrons are the most dangerous background particles. They can mimic a dark matter signal by inducing nuclear recoils in the detector. Since the cosmogenic neutron background is the dominant contribution, the use of an active Cherenkov veto is mandatory to reduce the rate of cosmogenic events to or below the level of the radiogenic one. The use of a large instrumented water tank volume can provide such a muon veto and allows, at the same time, to minimise the amount of further inner shielding layers (Pb or Cu), which in turn leads to a reduction of both the radiogenic neutron and $\gamma$ fluxes. 

Radiogenic background simulations show that the PE layer as innermost shielding is disadvantageous. Although the $\gamma$ background can be efficiently discriminated by using the double read-out channel approach, part of its low-energy fraction may potentially leak into the region of interest for dark matter search. In addition, the total rate in the detectors affects the dead-time of the experiment. Thus, also the minimisation of the $\gamma$ background is of importance. 
    
From these considerations, we conclude that options number 4 or 5 in Tab.~\ref{tab:DetailDesignOptions} best fulfill our requirements. Option 4 is the safer and thus preferred configuration, as it definitely leads to a reduced $\gamma$ rate and is less sensitive to the intrinsic radioactivity level of the stainless steel of the dry-well, which is yet to be determined.

As mentioned above, the actual background rate measured in our detectors will depend on the size, number, position and housing of the modules. By taking the example of shielding option 4, we briefly discuss the effect of using a segmented target made of an array of small crystals compared to the use of a single crystal of the same integrated mass. For this purpose, we ran two identical simulations: one using a single \unit[2.4]{kg} NaI cube, the other using 24 smaller \unit[100]{g} cubes, distributed over a larger volume. Looking at both cosmogenic and radiogenic neutrons, the segmented arrangement leads to an enhanced total rate for the same integrated detector mass, because of the larger surface to volume ratio. On the other hand, the segmentation allows for an anti-coincidence cut, removing events in which two or more detector modules trigger simultaneously. This cut is more efficient with respect to cosmogenic than radiogenic background, because in the first case more particles with higher energies enter the detector volume on average. In our simulation, we applied the anti-coincidence cut assuming a detection threshold of \unit[1]{keV}. In both arrangements the results yield similar nuclear recoil rates due to cosmogenic neutrons within statistical uncertainties. The obtained rates are of the order of $(3.5 \pm 0.7)\,$cts$\,$kg$^{-1}\,$yr$^{-1}$. On the other hand, the radiogenic neutron background is systematically slightly higher in the segmented detector setup, with a rate of $(5.5 \pm 0.1) \cdot 10^{-2}\,$cts$\,$kg$^{-1}\,$yr$^{-1}$ compared to $(4.4 \pm 0.1) \cdot 10^{-2}\,$cts$\,$kg$^{-1}\,$yr$^{-1}$. A more detailed study of the total estimated background rate will be the topic of a future publication once the final geometry of the detector volume will be assessed. However, the preliminary number of muon-induced neutron events already demonstrates the necessity to use a Cherenkov muon veto for COSINUS.

\section{Summary and conclusion}
In this work, we performed a conceptual design study of the experimental setup for the COSINUS experiment. We took into account all potential background sources external to the NaI detectors as well as a simplified geometry and compared different shielding configurations. The aim was to minimise the amount of ambient and cosmogenic background particles reaching the detector volume, while using the unavoidable intrinsic radiogenic background generated in the employed material as a benchmark. 

This study was used as the basis of the design of the COSINUS apparatus whose construction is foreseen to start in mid-2021.

Following the layout of option 4 in Tab.~\ref{tab:DetailDesignOptions}, the experimental setup will consist of a water tank with diameter and height of \unit[7]{m}, providing a shielding of at least \unit[3]{m} of water all around the detector volume where the NaI crystals will be inserted. The diameter of the dry-well has been set in order to host the \unit[8]{cm} layer of Cu around the cryostat, extending in height to line up with the inner Cu shield inside the cryostat, as shown in Fig.~\ref{fig:tank_simple}. To suppress the dominant cosmogenic neutron contribution, the water tank will be instrumented with PMTs to serve as an active Cherenkov veto. 

A publication describing the perfomance and the geometry of the muon veto is currently in preparation.

\begin{acknowledgements}
This work has been supported by the Austrian Science Fund FWF, Doctoral program No. W1252-N27 and the stand-alone project AnaCONDa [P 33026-N], and by the Gran Sasso Science Institute GSSI.
We thank INFN and LNGS for their support to the COSINUS experiment.
\end{acknowledgements}

\bibliographystyle{epj}
\bibliography{biblio}   

\end{document}